\documentclass[aps,prd,twocolumn,amsmath,amssymb,floatfix,nofootinbib,superscriptaddress]{revtex4-1}

\newcommand{\nhat}{\hat{ \mathbf{n}}}

\usepackage{graphicx}
\usepackage{dcolumn}
\usepackage{bm}
\usepackage[usenames, dvipsnames]{color}
\usepackage[normalem]{ulem}
\usepackage{xcolor}
\usepackage{hyperref}
\usepackage[utf8]{inputenc}

\graphicspath{ {figures/} }

\usepackage{epstopdf}
\usepackage{tabularx}
\newcolumntype{C}{>{\centering\arraybackslash}X}
\newcolumntype{R}{>{\raggedleft\arraybackslash}X}

\newcommand{\be}{\begin{eqnarray}}
\newcommand{\non}{\nonumber \\}
\newcommand{\ee}{\end{eqnarray}}

\newcommand{\wj}[6]{\left(
                           \begin{array}{ccc}
        \! #1\! & #2\!  & #3\!  \\
        \! #4\! & #5\!  & #6\!
                           \end{array}
                   \right)}

\usepackage[export]{adjustbox}

\def\nhat{\hat{\mathbf{n}}}

\usepackage{suffix}
\usepackage{mathtools}

\DeclarePairedDelimiterX\MeijerM[3]{\lparen}{\rparen}%
{\begin{smallmatrix}#1 \\ #2\end{smallmatrix}\delimsize\vert\,#3}

\newcommand\MeijerG[8][]{%
  G^{\,#2,#3}_{#4,#5}\MeijerM[#1]{#6}{#7}{#8}}

\WithSuffix\newcommand\MeijerG*[7]{%
  G^{\,#1,#2}_{#3,#4}\MeijerM*{#5}{#6}{#7}}

\newcommand{\mnras}{Mon. Not. R. Astron. Soc.}
\newcommand{\jcap}{J. Cosm. Astropart. Phys.}

\newcommand{\dd}{{\rm d}}

\newcommand{\bc}{{b_{\rm bc}}}

\newcommand{\imperial}{Department of Physics, Imperial College London, Blackett Laboratory, Prince Consort Road, London SW7 2AZ, UK}

\newcommand{\jhu}{Department of Physics \& Astronomy, Johns Hopkins University, Baltimore, MD 21218, USA}

\newcommand{\perimeter}{Perimeter Institute for Theoretical Physics, 31 Caroline St N, Waterloo, ON N2L 2Y5, Canada}

\newcommand{\york}{Department of Physics and Astronomy, York University, Toronto, ON M3J 1P3, Canada}

\newcommand{\CITA}{Canadian Institute for Theoretical Astrophysics,
University of Toronto, Toronto, ON M5H 3H8 Canada}

\begin{document}

\title{Probing correlated compensated isocurvature perturbations using scale-dependent galaxy bias}

\author{Selim~C.~Hotinli}
\affiliation{\imperial}

\author{James~B.~Mertens}
\affiliation{\perimeter}
\affiliation{\york}
\affiliation{\CITA}

\author{Matthew~C.~Johnson}
\affiliation{\perimeter}
\affiliation{\york}

\author{Marc~Kamionkowski}
\affiliation{\jhu}

\date{\today}

\begin{abstract}
Compensated isocurvature perturbations (CIPs) are modulations of the relative baryon and dark matter density that leave the total matter density constant. The best current constraints from the primary cosmic microwave background (CMB) are consistent with CIPs some two orders of magnitude larger in amplitude than adiabatic perturbations, suggesting that there may be a huge gap in our knowledge of the early Universe.  However, it was recently suggested by Barreira~et.~al. that CIPs which are correlated with the primordial curvature perturbation, as arises in some versions of the curvaton model, lead to a new observable: scale dependent galaxy bias. Combining a galaxy survey with an unbiased tracer of the density field facilitates a measurement of the amplitude of correlated CIPs that is free from cosmic variance, the main limitation on constraints from the primary CMB. Among the most promising tracers to use for this purpose is the remote dipole field, reconstructed using the technique of kinetic Sunyaev Zel'dovich (kSZ) tomography. In this paper, we evaluate the detection significance on the amplitude of correlated CIPs possible with next-generation CMB and galaxy surveys using kSZ tomography. Our analysis includes all relativistic contributions to the observed galaxy number counts and allows for both CIPs and primordial non-Gaussianity, which also gives rise to a scale dependent galaxy bias. We find that kSZ tomography can probe CIPs of comparable amplitude to the adiabatic fluctuations, representing an improvement of over two orders of magnitude upon current constraints, and an order of magnitude over what will be possible using future CMB or  galaxy surveys alone.
\end{abstract}

\maketitle

\section{Introduction}

\vspace*{-0.18cm}

Measurements of the cosmic microwave background (CMB) provide the bedrock for the standard cosmological model, $\Lambda$CDM. A central feature of $\Lambda$CDM is that perturbations are adiabatic, with inhomogeneities in dark matter, baryons, neutrinos, and photons all uniquely determined by the primordial curvature perturbations. Theories of the early Universe which have one degree of freedom, such as single field inflation, naturally predict purely adiabatic fluctuations. More generally, theories with multiple degrees of freedom can source isocurvature (entropy) perturbations, where the relative mixture of dark matter, baryons, neutrinos, and photons become independent degrees of freedom. While most forms of isocurvature perturbations are tightly constrained by existing measurements of the CMB~\cite{Akrami:2018odb}, there is a notable exception: compensated isocurvature perturbations (CIPs). CIPs are fluctuations of baryons and cold dark matter that leave the total matter perturbations unchanged and adiabatic. CIPs leave an imprint on the CMB only through terms that appear at second order in the matter density contrast, making them challenging to constrain~\citep{Holder:2009gd,Gordon:2009wx,Grin:2011tf,Grin:2011nk,Smith:2017ndr, Munoz:2015fdv,Grin:2013uya}. Current measurements from Planck~\cite{Akrami:2018odb} allow for an amplitude of CIPs roughly $580$ times larger than the amplitude of the adiabatic modes!~\footnote{More recently, constraints on CIPs from their effect on baryon acoustic oscillations~\citep{Heinrich:2019sxl} (BAO) were analyzed. It was shown that constraints comparable to those from the CMB are possible with future galaxy surveys.} This is a surprisingly large gap in our knowledge of the early Universe. A detection of CIPs can provide insight into both the number of primordial fields that contribute to the observed density fluctuations, as well as their decay channels~\cite{Lyth:2001nq,Moroi:2002rd,Lyth:2003ip}, strongly motivating new ways of searching for CIPs. 

Variations of the ratio between baryons and cold dark matter changes how structure is distributed in the Universe, altering how galaxies trace the total matter density~\cite{2011MNRAS.415.3113B,Schmidt:2016coo,Barreira:2019qdl}. This leads to a spatially varying galaxy-bias that relates the observed galaxy over-density to the total matter over-density. In particular, CIPs that are correlated with the primordial curvature perturbation (as arises in e.g. curvaton scenarios~\cite{Lyth:2001nq}) will introduce a scale-dependent galaxy-bias~\cite{Barreira:2019qdl},
similar to the effect of local-type primordial non-Gaussianity~\citep{Dalal:2007cu}. Because correlated CIPs induce a scale dependent galaxy bias, given an unbiased tracer of the total matter density, it is possible to use sample variance cancellation~\cite{McDonald:2008sh,Seljak:2008xr} to measure the amplitude of CIPs without cosmic variance, as suggested in Ref.~\cite{Barreira:2019qdl}. While it is possible to use different populations within a galaxy survey itself to measure scale-dependent bias, sample variance cancellation is in principle more powerful when using the technique of kSZ tomography~\cite{Munchmeyer:2018eey,Contreras:2019bxy}. The primary goal of this paper is to explore the potential for kSZ tomography to probe CIPs using future CMB and galaxy surveys. 

The technique of kSZ tomography~\cite{Zhang10d,Zhang:2015uta,Terrana:2016xvc,Deutsch:2017ybc,Smith:2018bpn} uses the correlation between redshift-binned galaxy number counts and the small-angular scale kSZ contribution to the CMB to reconstruct the three dimensional remote dipole field, the CMB dipole as observed at different locations in our Universe. The remote dipole field, which at any location is dominated by the Doppler effect associated with radial peculiar velocities, can be reconstructed with high fidelity on large angular scales using future surveys such as Simons Observatory~\citep{Ade:2018sbj} or CMB-S4~\citep{Abazajian:2016yjj} and LSST~\citep{Abell:2009aa} or DESI~\citep{Aghamousa:2016zmz}. The reconstruction is in principle of such high quality that it is superior to direct measurements of the density field from the galaxy survey itself, making kSZ tomography a powerful probe of inhomogeneities on the largest scales. These measurements can facilitate strong constraints on primordial non-Gaussianity~\cite{Munchmeyer:2018eey,Contreras:2019bxy}, the physics of cosmic acceleration~\cite{Pan:2019dax}, and inflationary cosmology~\cite{Cayuso:2019hen,Zhang:2015uta}.

Previous work~\cite{Munchmeyer:2018eey,Contreras:2019bxy} has found that future experiments will be able to detect local-type non-Gaussianity of order $\sigma_{f_{\rm NL}}\sim \mathcal{O}(1)$ by utilizing sample variance cancellation between the reconstructed remote dipole field and galaxy number counts. Depending on  assumptions, priors on various bias parameters, and whether internal sample variance cancellation is employed, this can represent up to an order of magnitude improvement on what is possible using the galaxy survey alone. Below, we find a similar level of improvement on the amplitude of CIPs when utilizing kSZ tomography. In particular, it will be possible to probe CIPs comparable in amplitude to the adiabatic perturbations, which can be thought of as a well-motivated target for future measurements.

The plan of the paper is as follows. In Sec.~\ref{sec:CIPs} we describe potential sources of and observable consequences of CIPs. In Sec.~\ref{sec:ksztomography} we review kSZ tomography, and then examine how well future surveys can measure correlated CIPs in Sec~\ref{sec:forecast}. We conclude in Sec.~\ref{sec:discussion}.

\section{CIPs and their observable consequences}\label{sec:CIPs}

In the early Universe, standard single-field inflation produces purely adiabatic curvature perturbations. If the fluctuations seeded in the early Universe were sourced by multiple fields, however, some fraction of these may be entropic (or isocurvature) perturbations where the fractional densities of baryons or dark matter vary with respect to radiation.
Isocurvature perturbations can be parameterized by a quantity
$S_{i\gamma}$, with $\gamma$ for photons and $i=\{b,c,\nu\}$ for baryons, cold-dark-matter (CDM) and neutrinos respectively, and 
\be
S_{i\gamma}=\frac{\delta n_i}{n_i}-\frac{\delta n_\gamma}{n_\gamma}\,,
\ee
where $n$ and $\delta n$ are 
the mean number density of a species and its fluctuations, respectively. 

The compensated isocurvature perturbations between baryons and CDM studied in this paper is a particular combination of these perturbations which leaves the total matter density fluctuations unchanged, where the baryon number density fluctuations are exactly compensated by those of CDM. We will define the compensated isocurvature mode with $\Delta$ as in the literature. The baryon and CDM isocurvature perturbations are then defined as 
\be
S_{b\gamma}=\Delta,\,\,\,\,S_{c\gamma}=-\frac{\rho_b}{\rho_c}\Delta\,
\ee
where $\rho_i$ is the energy density of species $i$. 
    
Compensated isocurvature perturbations may be sourced, for example, by a spectator scalar field that is subdominant in the early Universe with respect to the inflaton field driving the inflationary dynamics~\citep{He:2015msa}. In this scenario, after inflation ends, the inflaton decays into relativistic particles and its energy density scales like radiation, while the spectator field (curvaton) oscillates around its potential minimum, its energy density scaling like matter e.g.~\citep{Linde:1996gt,Lyth:2001nq,Lyth:2003ip,Moroi:2001ct,Lyth:2001nq,Moroi:2002rd,Gordon:2002gv}. Depending on the duration of this era, the curvaton may contribute significantly to curvature fluctuations of the Universe upon its decay into relativistic particles. 
    
If the curvaton decays into baryon number and CDM and also dominates the energy density of the Universe at its decay, the CIPs will be fully correlated with the adiabatic curvature fluctuations $\zeta$, satisfying 
\be\label{eq:CIPs_zeta}
\Delta=A\zeta\,,
\ee 
while any residual isocurvature perturbations other than CIPs that are uncorrelated with the adiabatic curvature fluctuations are well constrained by the CMB observations~\citep{Akrami:2018odb}. Similar to earlier works in the literature,~e.g.~\citep{He:2015msa,Heinrich:2016gqe,Heinrich:2019sxl}, we will focus on these ``correlated CIPs" and evaluate the detection significance of the amplitude $A$ below. {The two distinct curvaton decay scenarios that produce observationally relevant CIP amplitudes are either $A\simeq16$, if baryon (CDM) number is produced by (before) curvaton decay; or $A=-3$, if CDM (baryon) number is produced by (before) curvaton decay. Furthermore, in the former curvaton decay model where $A\simeq16$, the local non-Gaussianity is found to be relatively large, $f_{\rm NL}\simeq6$~\citep{Sasaki:2006kq,He:2015msa}, suggesting future experiments may disfavor the scenarios where CDM preceded the decay of curvaton. Note that an unambiguous statement along these lines will require constraining both the CIP amplitude and $f_{\rm NL}$ simultaneously, as we discuss in~Sec.~\ref{sec:forecast}.} 
    
In the absence of primordial isocurvature perturbations after recombination, baryons and CDM can be approximated to move together as a single fluid on large scales where non-gravitational forces can be neglected. However, both before recombination and in the presence of primordial isocurvature perturbations, there can be important differences in the distribution of baryons and CDM. For example, before recombination baryons are tightly coupled to photons while CDM is not. This leads to a modulation in the relative fraction of baryons and CDM on large scales while keeping the total matter density fixed, and therefore is a source of CIPs~\cite{2011MNRAS.415.3113B,Schmidt:2016coo,Barreira:2019qdl}. In addition, we may have the primordially sourced CIPs discussed above. As we will see shortly, primordial correlated CIPs can be distinguished from these more mundane sources of CIPs by their characteristic scale dependent imprint on the distribution of galaxies.

There are a few potential imprints of CIPs on the observed galaxy distribution. First, the sound horizon becomes spacetime dependent, altering the BAO feature in different regions of the Universe~\cite{Heinrich:2019sxl}. Second, modulating the density of baryons can modulate the strength of various feedback effects in the formation and evolution of galaxies. Finally, because only dark matter can cluster efficiently prior to recombination, modulating the density of dark matter will lead to a modulation in the growth of structure. It is this last effect that provides the dominant contribution on large scales, and which we focus on. 

As shown in Ref.~\cite{Barreira:2019qdl}, the leading effect of CIPs on galaxy density perturbations can be folded into a linear bias $b_{\rm bc}(z)$: 
\be\label{eq:galaxy_bias}
\delta_g(\mathbf{k},\tau)\simeq b(z)\,\delta_m(\mathbf{k},\tau)+\bc(z)\left[\delta_{\rm bc}(\mathbf{k},\tau)+f\Delta(\mathbf{k})\right],\non
\ee
where we have allowed for both pre-recombination CIPs $\delta_{\rm bc}$, as well as primordially sourced correlated CIPs, $f\equiv 1+{{\Omega}_b}/{{\Omega}_c}$, and we can relate $\Delta(\mathbf{k})$ to the total density perturbation by:
\be
\Delta = \frac{5 H^2 \Omega_m}{2 a k^2} A \ \delta_m.
\ee
Therefore, we see that primordially sourced correlated CIPs lead to a scale-dependent galaxy bias, becoming increasingly important on the largest scales. This scale dependence can be contrasted with the imprint of $\delta_{\rm bc}$, which is expected to be very small on scales larger than the BAO feature~\cite{Barreira:2019qdl}. Indeed, on the scales of interest ($\sim$ Gpc), $\delta_{\rm bc}$ is many orders of magnitude smaller than $\delta_m$ and can be safely neglected.

The bias $b_{\rm bc}(z)$ can be estimated in the separate Universe approximation by simply computing the effect of changing the baryon-CDM fraction on the number density of galaxies. We define
\begin{equation}
\bc(z) = \int dm \ n(m,z) \bc(m,z) \frac{\langle N(m) \rangle}{\bar{n}_g}\,,
\end{equation}
where $n(m,z)$ is the halo mass function, $\langle N(m) \rangle$ is the average number of galaxies per halo of mass $m$, $\bar{n}_g$ is the comoving number density of galaxies at fixed redshift, and
\begin{equation}
\bc (m,z) = \frac{1}{\delta_{\rm bc}} \left[ \frac{\tilde{n} (m,z)}{n(m,z)} -1 \right]\,,
\end{equation}
with
\begin{equation}
\delta_{\rm bc} =  \left(1+\frac{\Omega_b}{\Omega_c} \right) \Delta_b\,,
\end{equation}
and the mass function $\tilde{n}$ is evaluated with parameters:
\begin{equation}
\tilde{\Omega}_b = \left(1+\Delta_b \right) \Omega_b, \ \ \ \tilde{\Omega}_c = \left(1 - \frac{\Omega_b}{\Omega_c} \Delta_b \right) \Omega_c\,.
\end{equation}
To evaluate $\bc(z)$, we use the mass function and Halo Occupation Distribution (HOD) model for $\langle N(m) \rangle$ and $\bar{n}_g$ described in Ref.~\cite{Smith:2018bpn}. For parameters consistent with the LSST gold sample used in the forecast below, we find that a quadratic polynomial provides a good fit over the relevant range of redshifts:
\begin{equation}
\label{eq:bcb}
\bc (z) \simeq -(0.16 + 0.2z +0.083z^2)\,.
\end{equation}

\begin{figure}[t]
    \centering
    \includegraphics[width = \columnwidth]{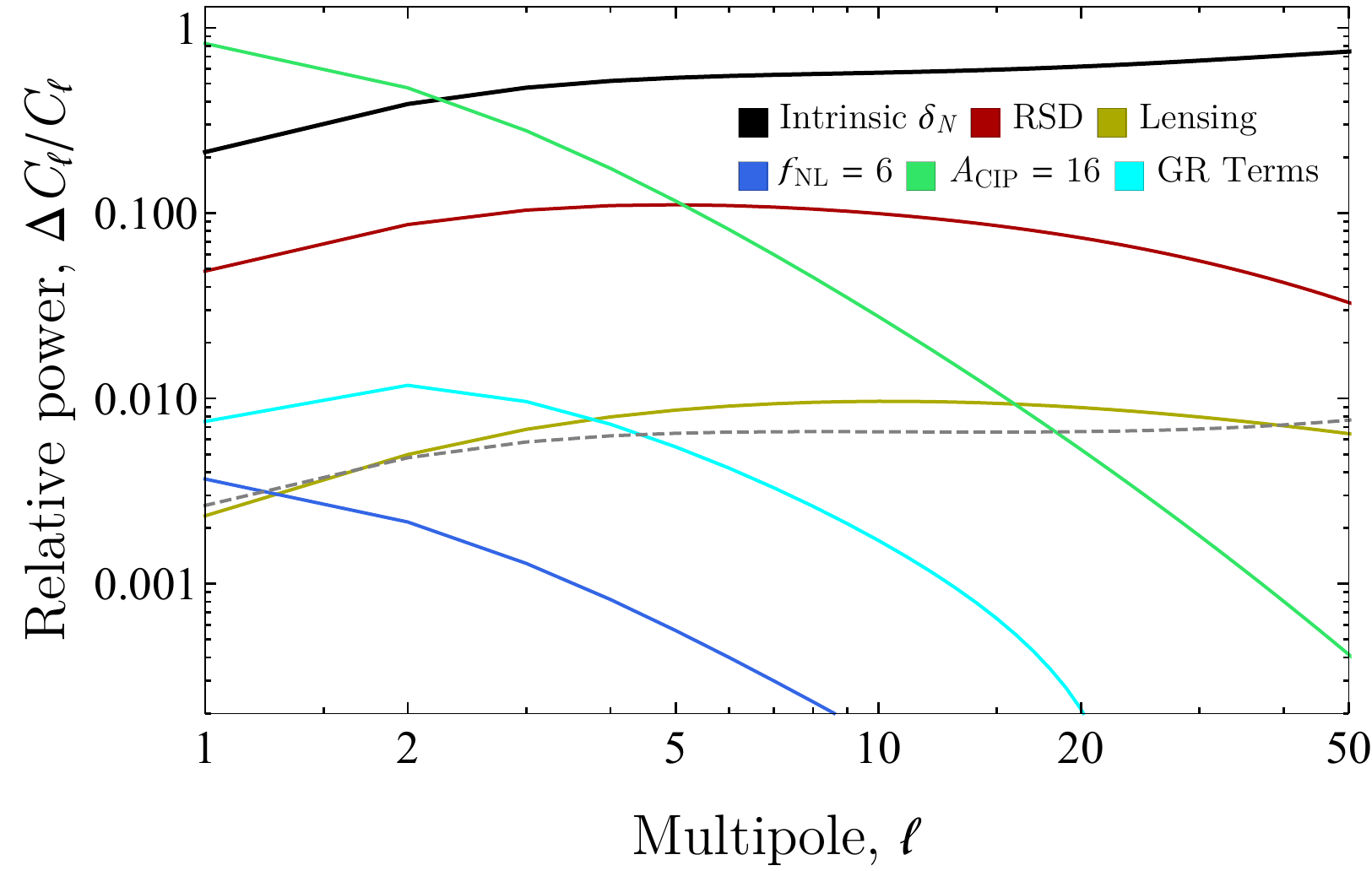}
    \caption{
    Relative contributions to the angular galaxy number counts power spectrum, as labeled in the figure, in a tophat redshift bin from redshift $z = 1.6$ to $z=1.7$. Shot noise from a galaxy survey is shown in dashed gray. }
    \label{fig:pow_specs}
\end{figure}

The total observed galaxy number counts receive contributions not only from CIPs and intrinsic density perturbations (D), but from all linear-order general relativistic and lightcone projection effects, including redshift space distortions (RSDs), lensing (L), and additional relativistic contributions (GR) that are important on large scales
\cite{2011PhRvD..84f3505B,2011PhRvD..84d3516C}. The spectrum of the total observed galaxy number counts
\begin{equation}
C_\ell^{N,N} = 4\pi \int \frac{dk}{k} \mathcal{P}(k) |\Delta_{\ell}^{\rm N}(k)|^2
\end{equation}
is defined by the transfer function
\begin{equation}
  \Delta_{\ell}^{\rm N}(k) = \Delta_{\ell}^{\rm D}(k) +  \Delta_{\ell}^{\rm RSD}(k) + \Delta_{\ell}^{\rm L}(k) +
  \Delta_{\ell}^{\rm GR}(k)\,.
\end{equation}
The power spectrum is defined by $\mathcal{P}(k)=A_s (k/k_0)^{n_s - 1}$, and
the transfer function for the intrinsic galaxy perturbations in a redshift bin is explicitly given by
\begin{align}
\Delta_{\ell}^{\rm D}(k) & =\int d\chi \tilde{W}(\chi)\,\bigg[ \frac{5f}{3} A\,b_{\rm cb}(z) S_{\psi}(k,\chi) + (b_{\rm G}(z) \nonumber \\
 & -b_{\rm A}(z)/3 + b_{\rm NG}(z) )\,S_{\delta_M,{\rm syn}}(k,\chi) \bigg]\,j_{\ell}(k\chi),
\end{align}
with $\tilde{W}(\chi)$ a window function selecting the relevant redshift bin, $S_{\delta_M,{\rm syn}}$ the time-evolution function for cold dark matter in comoving-synchronous gauge, and $S_{\psi}$ the time-evolution function for the Newtonian potential. Galaxy bias ($b_{\rm G}$) and alignment bias~\cite{0903.4929} ($b_{\rm A}$) are marginalized over, and non-Gaussianities are also modeled as an effective scale-dependent bias, $b_{\rm NG} \propto f_{\rm NL}$. These bias functions, as well as the remaining contributions to the number counts transfer function, are modeled identically to \cite{Contreras:2019bxy}. We show the relative contributions from each of these effects to the total power in Fig~\ref{fig:pow_specs}.

\section{kSZ tomography}\label{sec:ksztomography}

The kinetic Sunyaev Zel'dovich (kSZ) effect, Thomson scattering of CMB photons from free electrons in the late Universe, provides the dominant source of temperature anisotropies on small angular scales (corresponding to $\ell \agt 4000$). The temperature anisotropy induced by the kSZ effect in the $\nhat$ direction is
\be
\frac{T(\nhat)}{T_{\rm CMB}}\Big|_{\rm kSZ} = -\sigma_T \int\dd\chi \ a(\chi)\,n_e(\chi\nhat)\,v_{\rm eff}(\chi\nhat),
\ee
where $T_{\rm CMB}$ is the mean CMB temperature, $\chi$ is the comoving distance, $\sigma_T$ is the Thomson cross section, $a$ is the scale factor, $n_e$ is the free electron number density, and $v_{\rm eff}=3\int\dd^2\nhat_e\Theta_1(\nhat,\nhat_e)\nhat\cdot\nhat_e/(4\pi)$ is the remote CMB dipole field projected along the line of sight.
On small scales, the remote dipole field can be approximated by the Doppler term induced by Newtonian peculiar velocities, $v_{\rm eff} \simeq \vec{v}_{\rm pec}\cdot\nhat$. However, to probe the large scales we consider here it is important to include the contributions from the Sachs Wolfe, Integrated Sachs Wolfe, and primordial Doppler effects. A complete description of the contributions to the remote dipole field can be found in Refs.~\cite{Terrana:2016xvc,Deutsch:2017ybc}. Most of the cosmological information is contained in $v_{\rm eff}$, while $n_e$ depends primarily on astrophysics and non-linear large scale structure; see Ref.~\cite{Smith:2018bpn} for a detailed discussion of this point.

Kinetic Sunyaev Zel'dovich tomography aims to extract the cosmological information from the kSZ effect by using measurements of the CMB and a tracer of the electron density, such as a galaxy survey, to reconstruct the remote dipole field. The reconstructed dipole field, in cross-correlation with the galaxy survey or primary CMB, can then be used to estimate cosmological parameters. In the present context, it is important to note that the remote dipole field is an unbiased tracer of the total density. Cross-correlation with a galaxy survey can therefore take full advantage of sample variance cancellation in order to extract (scale dependent) galaxy bias to high precision.

More specifically, we can write a quadratic estimator for the remote dipole field averaged in a set of tophat redshift bins labeled by index $\alpha$ as:
\begin{equation}
\label{eq:estimator1}
\begin{split}
\widehat{v}_{{\rm eff}, \ell m}^{\alpha}&  = \,\, b_v^\alpha {N^{vv}_{\alpha \ell}} \\ 
 & \times\!\!\!\!\!\!\!\!\sum_{\ell_1m_1\ell_2m_2}\!\!\!\!\!\!(-1)^m \
\Gamma_{\ell_1\ell_2\ell}^\alpha
\wj{\ell_1}{\ell_2}{\ell}{m_1}{m_2}{-m} 
\frac{a^{T}_{\ell_1m_1} \delta^{\alpha}_{g,\ell_2m_2}}{C^{TT}_{\ell_1} C^{gg}_{\alpha \ell_2} },  
\end{split}
\end{equation}
where
\begin{equation}
\Gamma_{\ell_1\ell_2 \ell}^ \alpha = \sqrt{\frac{(2\ell_1+1)(2\ell_2+1)(2\ell+1)}{4\pi}} \wj{\ell_1}{\ell_2}{\ell}{0}{0}{0} \ C^{\tau g}_{\alpha,\ell_2}\,,
\end{equation}
and the reconstruction noise (e.g. variance of the estimator) is defined by
\begin{equation}
\label{eq:estimator1noise}
\frac{1}{N^{vv}_{\alpha \ell}} = \frac{1}{(2\ell+1)} \sum_{\ell_1\ell_2}
\frac{\Gamma_{\ell_1\ell_2\ell}^\alpha \ \Gamma_{\ell_1\ell_2\ell}^\alpha}{C^{TT}_{\ell_1} C^{gg}_{\alpha \ell_2} }. 
\end{equation}
In these expressions, $C^{TT}_{\ell_1}$ is the measured CMB temperature power spectrum, $C^{gg}_{\alpha \ell_2}$ is the measured spectrum of the galaxy number counts in each bin, and $C^{\tau g}_{\alpha,\ell_2}$ is the cross-power of the optical depth and galaxy number counts in each bin. In the absence of an external tracer of the electron distribution~\cite{Madhavacheril:2019buy}, there is in principle a significant model uncertainty in $C^{\tau g}_{\alpha,\ell_2}$. This uncertainty manifests itself as a multiplicative ``optical depth bias" $b_v^\alpha$ on the reconstructed dipole field which must be marginalized over in any cosmological analysis (see e.g.  Refs.~\cite{Mueller:2014nsa,Battaglia2016,Smith:2018bpn,Madhavacheril:2019buy} for further discussion). The reconstruction noise can in principle become arbitrarily small in the limit where the CMB and number counts can be probed on arbitrarily small angular scales. In reality, the reconstruction noise is limited by the instrumental noise of the CMB experiment and shot noise of the galaxy survey, since this places an effective upper limit in $\ell$ on the sum in Eq.~\ref{eq:estimator1noise}. The expected bin-averaged dipole field signal is computed as in Ref.~\cite{Deutsch:2017ybc}.

\section{Forecasts}\label{sec:forecast}

\begin{table*}[htb]
\renewcommand{\arraystretch}{1.75}
\begin{tabularx}{0.99\textwidth}{rCCCCCCCCCCCCCC}
Parameter & $A$ & $10^{9}A_{s}$ & $n_{s}$ & $\Omega_{b}$ & $\Omega_{c}$ & $h$ & $\tau$ & $b_{v}(z)$ & $b_{\rm G}(z)$ & $b_{\rm A}(z)$ & $f_{\rm evo}(z)$ & $s(z)$ & $b_{\rm bc}(z)$ \\
\hline
Fiducial value  & 0 & 2.2 & 0.96 & 0.0528 & 0.2647 & 0.675 & 0.06 & 1 & $\dagger$ & 0 & $\dagger$ & $\dagger$ & Eq.~\ref{eq:bcb} \\
\end{tabularx}
\caption{Various parameters, bias functions, and their
fiducial values. The biases $b_v,\, b_G,\, b_A,\, f_{\rm evo},\, s$, that we refer to
throughout are, respectively, the optical depth bias, the galaxy bias, the
alignment bias, the evolution bias, and the magnification bias. The fiducial values of bias functions indicated with a $\dagger$ vary
with redshift, the modeling of which is described in \cite{Contreras:2019bxy}.
}
\label{table:params_list}
\end{table*}

\begin{figure*}[htb]
    \centering
    \includegraphics[width = 1.0\textwidth]{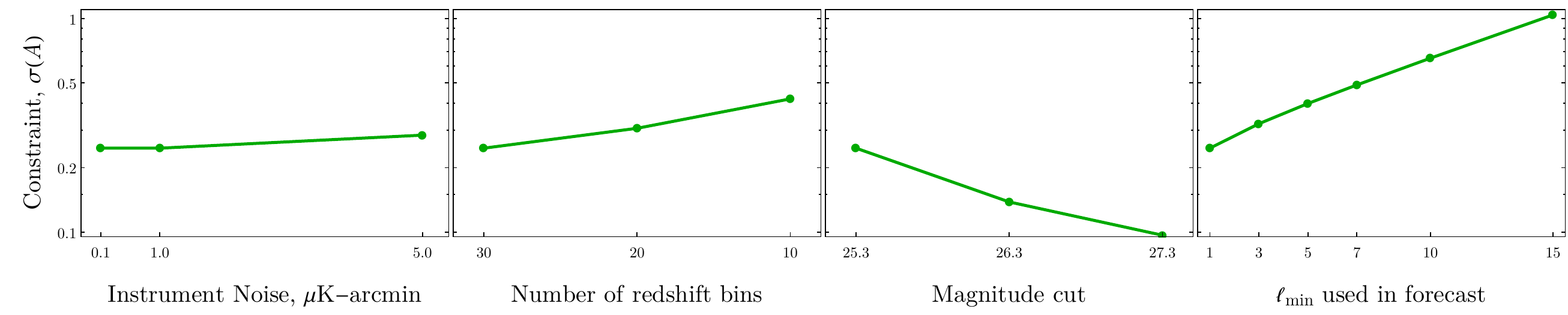}
    \caption{The impact of changing various parameters relevant for, or related to, experiments for the ``fiducial'' forecast we perform.}
    \label{fig:vary_exp_params}
\end{figure*}

We now examine how well future experiments will be able to measure $A$, assuming an LSST-``gold sample"-like galaxy survey, and kSZ reconstruction from a CMB-S4-like survey. We follow the prescription used in~\cite{Contreras:2019bxy} 
in order to compute galaxy number densities, the kSZ remote dipole field, and the corresponding noise for each tracer. The galaxy number densities follow from earlier work,~e.g.~\cite{2011PhRvD..84d3516C,2011PhRvD..84f3505B,Lorenz:2017iez,Alonso:2015uua},
and the kSZ signal from~e.g.~\cite{Deutsch:2017ybc}. 
We make use of information from each of these tracers individually, as well as the cross-correlations. The Fisher matrix we compute thus has the form
\begin{equation}
\label{eq:fisherM}
F_{\alpha\beta} = \sum_{\ell=\ell_{\rm min}}^{\ell_{{\rm max}}}\frac{2\ell+1}{2}{\rm Tr}\left[\left(\partial_{\alpha}\mathbf{C}_{\ell}\right)\mathbf{C}_{\ell}^{-1}\left(\partial_{\beta}\mathbf{C}_{\ell}\right)\mathbf{C}_{\ell}^{-1}\right] + F^{\rm CMB}_{\alpha\beta}\,,
\end{equation}
where the covariance matrix $\mathbf{C_\ell}$ is given by
\begin{equation}
\label{eq:covM}
\mathbf{C}_{\ell}=\left(\begin{array}{cc}
C_{\ell}^{{\rm N,N}} & C_{\ell}^{{\rm N,kSZ}} \\
C_{\ell}^{{\rm kSZ,N}} & C_{\ell}^{{\rm kSZ,kSZ}} \\
\end{array}\right) + N_{\ell}\,.
\end{equation}
The individual contributions to the covariance matrix are the spectra $C_{\ell}^{{\rm X,Y}}$, where $X,Y \in \{{\rm N, kSZ}\}$, and are the angular power spectra and cross-spectra of the galaxy number counts and kSZ remote dipole field. The noise computed for each tracer is denoted by $N_\ell$. For the galaxy number counts, we assume the dominant source of noise is shot noise from an LSST-like survey. Calibration errors may also exist on large scales that we do not explicitly model \cite{Weaverdyck:2017ovf}, although we do explore the dependence of detection prospects on a maximum available $\ell$ in Fig.~\ref{fig:vary_exp_params}. For the kSZ reconstruction, the noise is the reconstruction noise given by Eq.~\ref{eq:estimator1noise}, which we assume is uncorrelated with the galaxy shot noise.  The CMB contribution to the Fisher matrix, $F^{\rm CMB}_{\alpha\beta}$, is computed using information from the lensed CMB temperature and polarization power spectra, and is not cross-correlated with the galaxy survey nor remote dipole field. This term acts as an effective prior on standard cosmological parameters only.
Lastly, we compute derivatives of the covariance matrices analytically for all cosmological parameters and bias functions, except for the cosmological parameters $\Omega_b$, $\Omega_c$, and $h$, which we compute numerically. We test for numerical convergence by varying all relevant numerical parameters.

For our fiducial results, we sum over $1\leq\ell\leq60$; the vast majority of constraining power on $A$ and $f_{\rm NL}$ comes from $\ell \lesssim 30$. We assume information from a galaxy survey is available in 30 (tophat) redshift bins from $z=0$ to $z=3$ (so $\sigma_z \lesssim 0.05$), and a magnitude limit corresponding to the LSST gold sample, $r_{\rm max} = 25.3$.
For reconstruction of the remote dipole field, we assume modes up to $\ell$ of 9000 are available for reconstruction, subject to a 1.0 $\mu$K-arcmin noise and 1 arcmin beam for the CMB experiment. We explore the implications of varying this noise, and do not find our constraints change substantially: most of our signal comes from the largest angular scales, where the remote dipole field reconstruction noise is sufficiently low even for a much larger instrument noise.

The main quantity we report is
$\sigma_\alpha = \sqrt{F^{-1}_{\alpha\alpha}}$. We marginalize over standard cosmological parameters, as well as different bias functions. The full list of cosmological parameters we marginalize over, as well as the bias functions, are described in Table~\ref{table:params_list}  unless stated otherwise.
We examine $\sigma_A$ as a function of different ingredients in the forecast, in order to assess how much additional constraining power is available once new probes are added and theoretical considerations modified.
The constraints we find on $A$ for our ``fiducial'' model described above, as well as for different combinations probes, are shown in Table~\ref{table:forecasts}.
Notably, the remote dipole field improves constraining power over galaxy number densities alone by over an order of magnitude. Fixing standard cosmological parameters and bias functions in addition does not considerably improve constraining power, however we do find a moderate degeneracy of $A$ with the effects of non-Gaussianity, such that additionally marginalizing over $f_{\rm NL}$ worsens constraining power by a factor of order 2. There is also a minor degeneracy with general relativistic and lightcone projection effects: although less important, we find a parameter bias of order $1.5\sigma$ in $A$ when these effects (excluding lensing) are not modeled, suggesting such effects should be properly accounted for when studying isocurvature perturbations using large-scale survey data.

\begin{table}[bt]
\centering
\renewcommand{\arraystretch}{1.25}
\begin{tabularx}{0.9\columnwidth}{rC}
 Forecast ingredients & $\sigma_{A}$ \\ 
 \hline
 N only & $3.8$ \\ 
 N + CMB & $3.2$ \\
{\bf N + CMB + kSZ} & $0.25$ \\
N + CMB + kSZ + fixed cosmology & $0.23$ \\
\hspace{0.5cm} N + CMB + kSZ + variable $f_{\rm NL}$ & $0.49$
\end{tabularx}
\caption{
\label{table:forecasts}
The fiducial uncertainty in $A$ from the model described in the text is bold. Lines above this exclude the cross- and auto-correlation with the kSZ remote dipole field, and additionally exclude the high-$\ell$ CMB prior on standard cosmological parameters. Lines below fix all cosmological parameters and bias functions, or additionally marginalize over $f_{\rm NL}$ with a fiducial value of zero. }
\end{table}

\begin{figure}[htb]
    \centering
    \includegraphics[width = \columnwidth]{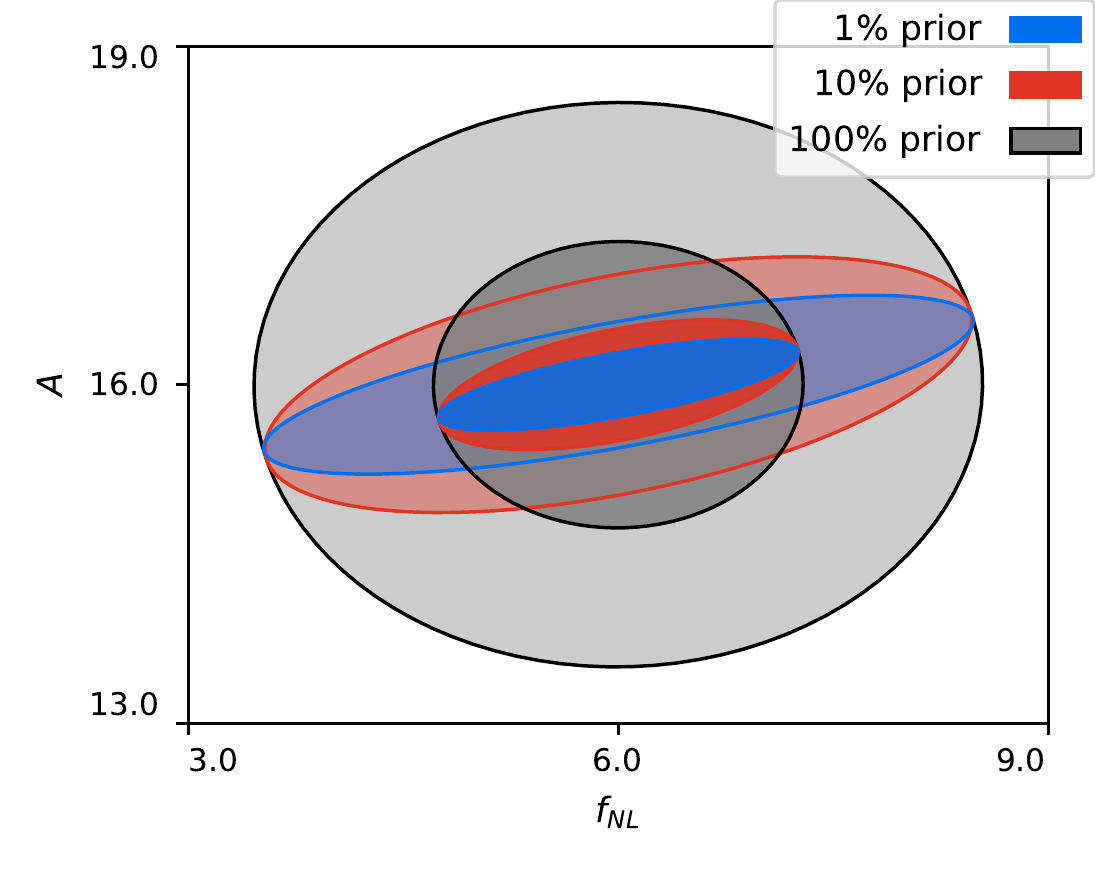}
    \caption{Parameter covariance between $A$ and $f_{\rm NL}$, given several choices for a prior on $\bc$.}
    \label{fig:AfNL_degeneracy}
\end{figure}

In order to check how robust the uncertainties we report are to the fiducial values we choose, as well as to draw a connection to a specific model, we re-evaluate our results for a value of $A=16$ and $f_{\rm NL} = 6$, corresponding to particular curvaton decay models. Because the fiducial value of $A$ is no longer zero, the bias function $\bc(z)$ should be marginalized over. Changes in this bias function are highly degenerate with changes in $A$, so we must place a prior on the function in order to obtain meaningful results. Enforcing a condition on the sign of $\bc$, or adding a ``100\%'' prior $\sigma(\bc) = \bc$ on the function in each redshift bin, results in an uncertainty in $A$ of $\sigma_A = 5.8$. The constraint scales down to $\sigma_A = 0.89$ for a 10\% prior, and $\sigma_A = 0.53$ for a 1\% prior, nearly recovering the results reported in Table~\ref{table:forecasts}. As primordial non-Gaussianity may be sourced through other additional mechanisms, we have marginalized over $f_{\rm NL}$ and $A$ separately. The degeneracy between these two parameters can be explicitly seen in Fig~\ref{fig:AfNL_degeneracy} as a function of the prior on $\bc$. Even with the weakest prior, we see that a definitive detection of this scenario can be made with future datasets.

We lastly show how the uncertainty $\sigma_A$ varies due to experimental parameters that have not been marginalized over. In particular,
we vary: the $\ell$ summed over in Eq.~\ref{eq:fisherM}, the number and width of the redshift bins we consider (which stand in as an effective redshift uncertainty), the galaxy survey magnitude limit, and CMB experiment noise. These results are summarized in Figure~\ref{fig:vary_exp_params}. The results generally do not change significantly as these are varied, with two exceptions. First, improving the magnitude limit from the LSST gold sample ($r=25.3$) to a less conservative cut ($r=27.3$) can improve things by another possible factor of order 2. Second, without a reliable survey or remote dipole field reconstruction on large angular scales, low-$\ell$ multipoles may not be accessible, degrading our constraint by a similar factor.

\section{Discussion}\label{sec:discussion}

In this paper, we have shown that by measuring the scale-dependent galaxy bias with the sample variance cancellation technique using kSZ tomography from upcoming CMB experiments and galaxy-surveys, constraining the correlated CIP amplitude at order one at high significance will be possible in the near future. We also considered a curvaton model of inflation and demonstrated that our method will be able to constrain the CIP amplitude and the local non-Gaussianity predicted by this model at high significance.  

As our understanding of the fundamental characteristics of the Universe advances, we might find it generally useful to know whether baryon and CDM fluctuations trace the total matter, or whether CIPs produced at early times correspond to a significant source of fluctuations in the Universe. From a phenomenological perspective, better constraints on the CIPs may rule out generic models of many field inflation, for example, or allow for less ambiguous measurements of early Universe signatures, such as primordial non-Gaussianity which may be degenerate with the CIPs. Constraints from the CMB measurements currently allow for CIPs to be up to a few orders of magnitude larger than the adiabatic fluctuations~\citep{Grin:2013uya}, while forecasts that use the upcoming CMB and galaxy surveys alone suggest it will still be hard to rule out scenarios where CIPs dominate over adiabatic fluctuations, or to distinguish between the different CIP scenarios discussed above, for example, with high significance. With an order of magnitude improvement on CIP constraints, here, we have been able to show that these issues may be resolved by measuring the galaxy-bias through sample-variance cancellation using the reconstructed remote dipole field from kSZ tomography.

In addition to the kSZ effect considered in this paper, one can also consider cross-correlating with other tracers of large-scale structure such as the remote quadrupole field from measurements of the polarized Sunyaev Zel'dovich effect e.g.~\citep{10.1093/mnras/190.3.413,Kamionkowski:1997na,Deutsch:2017cja,Deutsch:2017ybc,Meyers:2017rtf}, and the transverse velocity fields from the moving lens effect e.g.~\citep{1983Natur.302..315B,Hotinli:2018yyc,Yasini:2018rrl}.
Including these effects in our forecasts, we do not see a significant improvement upon the constraints presented in this work, although we note that using these effects without kSZ can still considerably improve upon past constraints. We leave considering additional probes of large-scale structure, such as galaxy-galaxy lensing, to future work. 

Our study focused on isocurvature perturbations modes that are correlated with the adiabatic fluctuations, as predicted by the curvaton models we consider. In principle, CIPs can be partially correlated (or uncorrelated) with the adiabatic perturbations. In relation to the galaxy-bias studied here, uncorrelated CIPs result in a halo over-density that is not fully correlated with matter over-density, inducing a so-called `stochastic' halo-bias on large scales~\citep{Baumann:2012bc}. The stochastic halo-bias can arise in many field models of inflation, for example, where the small-scale matter power may get redistributed in the presence of an additional field that do not contribute to the curvature fluctuations, and is not correlated with the gravitational potential. In these cases, the bias inferred from the cross-correlations of the halo over-density and matter over-density will differ from the bias inferred from halo auto-correlations, where the latter will see a boost compared to the former, which will be unaffected from uncorrelated fluctuations. The sample-variance cancellation method we use with the kSZ tomography will fail to detect the contribution from a stochastic contribution, as it utilizes the cross-correlations of the remote dipole field (an unbiased tracer of the matter over-density) and the galaxy over-density, in order to constrain the scale-dependent galaxy-bias. Moreover, any uncorrelated bias contributes as noise to this measurement, further worsening the significance of our constraints. It is thus hard to imagine taking advantage of the sample-variance cancellation in the case of uncorrelated CIPs. Nevertheless, depending on the scale dependence of the uncorrelated modes, it may still be possible to get competitive constraints on the CIP amplitude from measurement of the scale-dependent galaxy-bias using galaxy number-counts only, for example, compared to using CMB and BAO reconstruction alone, as can be seen from Table~\ref{table:forecasts}. We leave a more detailed study of the stochastic bias to future work.   

Lastly, we note that the current competitive studies of the scale-dependent galaxy-bias such as the one afforded by photometric quasar searches report stringent constraints on local non-Gaussianity, e.g.~$-49 < f_{\rm NL} <31$ \citep{Leistedt:2014zqa}, which can be translated into similar constraints on the CIP amplitude, $A$, by comparing the contribution to the transfer function of the intrinsic galaxy perturbations from local non-Gaussianity and the CIPs. We find that these contributions are similar at $\sim\mathcal{O}(1)$, suggesting that photometric quasar studies can already improve on current CMB constraints significantly. We leave a more careful analysis to an upcoming work. 

Advances in the precision of small-scale cosmology measurements from the near-future CMB and galaxy surveys will provide new opportunities to study the fundamental nature of the Universe on largest scales. We have used kSZ reconstruction and sample-variance cancellation in order to constrain correlated compensated isocurvature fluctuations on large scales and showed that our method will improve the detection significance by over an order of magnitude.

\section{Acknowledgements}

We would like to thank Colin Hill, Simone Ferraro, Mathew Madhavacheril, and Emmanuel Schaan for helpful discussions. SCH acknowledges the support of a visitor grant from the New-College Oxford/Johns-Hopkins Centre for Cosmological Studies, the Imperial College President’s Scholarship and the hospitality of the Perimeter Institute where part of this work was completed.
This research was supported in part by Perimeter Institute for
Theoretical Physics. Research at Perimeter Institute is supported by the Government of Canada through
the Department of Innovation, Science and Economic Development Canada and by the Province of Ontario
through the Ministry of Research, Innovation and Science. MCJ is supported by the National
Science and Engineering Research Council through a Discovery grant. JBM acknowledges support as a
CITA National Fellow.
MK was supported in part by NASA Grant No.\ NNX17AK38G, NSF Grant No.\ 1818899, and the Simons Foundation.
We acknowledge use of the cosmicfish package \cite{Raveri:2016leq}.

\bibliography{cip_constraints}

\begin{thebibliography}{56}%
\makeatletter
\providecommand \@ifxundefined [1]{%
 \@ifx{#1\undefined}
}%
\providecommand \@ifnum [1]{%
 \ifnum #1\expandafter \@firstoftwo
 \else \expandafter \@secondoftwo
 \fi
}%
\providecommand \@ifx [1]{%
 \ifx #1\expandafter \@firstoftwo
 \else \expandafter \@secondoftwo
 \fi
}%
\providecommand \natexlab [1]{#1}%
\providecommand \enquote  [1]{``#1''}%
\providecommand \bibnamefont  [1]{#1}%
\providecommand \bibfnamefont [1]{#1}%
\providecommand \citenamefont [1]{#1}%
\providecommand \href@noop [0]{\@secondoftwo}%
\providecommand \href [0]{\begingroup \@sanitize@url \@href}%
\providecommand \@href[1]{\@@startlink{#1}\@@href}%
\providecommand \@@href[1]{\endgroup#1\@@endlink}%
\providecommand \@sanitize@url [0]{\catcode `\\12\catcode `\$12\catcode
  `\&12\catcode `\#12\catcode `\^12\catcode `\_12\catcode `\%12\relax}%
\providecommand \@@startlink[1]{}%
\providecommand \@@endlink[0]{}%
\providecommand \url  [0]{\begingroup\@sanitize@url \@url }%
\providecommand \@url [1]{\endgroup\@href {#1}{\urlprefix }}%
\providecommand \urlprefix  [0]{URL }%
\providecommand \Eprint [0]{\href }%
\providecommand \doibase [0]{http://dx.doi.org/}%
\providecommand \selectlanguage [0]{\@gobble}%
\providecommand \bibinfo  [0]{\@secondoftwo}%
\providecommand \bibfield  [0]{\@secondoftwo}%
\providecommand \translation [1]{[#1]}%
\providecommand \BibitemOpen [0]{}%
\providecommand \bibitemStop [0]{}%
\providecommand \bibitemNoStop [0]{.\EOS\space}%
\providecommand \EOS [0]{\spacefactor3000\relax}%
\providecommand \BibitemShut  [1]{\csname bibitem#1\endcsname}%
\let\auto@bib@innerbib\@empty
\bibitem [{\citenamefont {Akrami}\ \emph {et~al.}(2018)\citenamefont {Akrami}
  \emph {et~al.}}]{Akrami:2018odb}%
  \BibitemOpen
  \bibfield  {author} {\bibinfo {author} {\bibfnamefont {Y.}~\bibnamefont
  {Akrami}} \emph {et~al.} (\bibinfo {collaboration} {Planck}),\ }\href@noop {}
  {\  (\bibinfo {year} {2018})},\ \Eprint {http://arxiv.org/abs/1807.06211}
  {arXiv:1807.06211 [astro-ph.CO]} \BibitemShut {NoStop}%
\bibitem [{\citenamefont {Holder}\ \emph {et~al.}(2010)\citenamefont {Holder},
  \citenamefont {Nollett},\ and\ \citenamefont {van Engelen}}]{Holder:2009gd}%
  \BibitemOpen
  \bibfield  {author} {\bibinfo {author} {\bibfnamefont {G.~P.}\ \bibnamefont
  {Holder}}, \bibinfo {author} {\bibfnamefont {K.~M.}\ \bibnamefont {Nollett}},
  \ and\ \bibinfo {author} {\bibfnamefont {A.}~\bibnamefont {van Engelen}},\
  }\href {\doibase 10.1088/0004-637X/716/2/907} {\bibfield  {journal} {\bibinfo
   {journal} {Astrophys. J.}\ }\textbf {\bibinfo {volume} {716}},\ \bibinfo
  {pages} {907} (\bibinfo {year} {2010})},\ \Eprint
  {http://arxiv.org/abs/0907.3919} {arXiv:0907.3919 [astro-ph.CO]} \BibitemShut
  {NoStop}%
\bibitem [{\citenamefont {Gordon}\ and\ \citenamefont
  {Pritchard}(2009)}]{Gordon:2009wx}%
  \BibitemOpen
  \bibfield  {author} {\bibinfo {author} {\bibfnamefont {C.}~\bibnamefont
  {Gordon}}\ and\ \bibinfo {author} {\bibfnamefont {J.~R.}\ \bibnamefont
  {Pritchard}},\ }\href {\doibase 10.1103/PhysRevD.80.063535} {\bibfield
  {journal} {\bibinfo  {journal} {Phys. Rev.}\ }\textbf {\bibinfo {volume}
  {D80}},\ \bibinfo {pages} {063535} (\bibinfo {year} {2009})},\ \Eprint
  {http://arxiv.org/abs/0907.5400} {arXiv:0907.5400 [astro-ph.CO]} \BibitemShut
  {NoStop}%
\bibitem [{\citenamefont {Grin}\ \emph
  {et~al.}(2011{\natexlab{a}})\citenamefont {Grin}, \citenamefont {Dore},\ and\
  \citenamefont {Kamionkowski}}]{Grin:2011tf}%
  \BibitemOpen
  \bibfield  {author} {\bibinfo {author} {\bibfnamefont {D.}~\bibnamefont
  {Grin}}, \bibinfo {author} {\bibfnamefont {O.}~\bibnamefont {Dore}}, \ and\
  \bibinfo {author} {\bibfnamefont {M.}~\bibnamefont {Kamionkowski}},\ }\href
  {\doibase 10.1103/PhysRevD.84.123003} {\bibfield  {journal} {\bibinfo
  {journal} {Phys. Rev.}\ }\textbf {\bibinfo {volume} {D84}},\ \bibinfo {pages}
  {123003} (\bibinfo {year} {2011}{\natexlab{a}})},\ \Eprint
  {http://arxiv.org/abs/1107.5047} {arXiv:1107.5047 [astro-ph.CO]} \BibitemShut
  {NoStop}%
\bibitem [{\citenamefont {Grin}\ \emph
  {et~al.}(2011{\natexlab{b}})\citenamefont {Grin}, \citenamefont {Dore},\ and\
  \citenamefont {Kamionkowski}}]{Grin:2011nk}%
  \BibitemOpen
  \bibfield  {author} {\bibinfo {author} {\bibfnamefont {D.}~\bibnamefont
  {Grin}}, \bibinfo {author} {\bibfnamefont {O.}~\bibnamefont {Dore}}, \ and\
  \bibinfo {author} {\bibfnamefont {M.}~\bibnamefont {Kamionkowski}},\ }\href
  {\doibase 10.1103/PhysRevLett.107.261301} {\bibfield  {journal} {\bibinfo
  {journal} {Phys. Rev. Lett.}\ }\textbf {\bibinfo {volume} {107}},\ \bibinfo
  {pages} {261301} (\bibinfo {year} {2011}{\natexlab{b}})},\ \Eprint
  {http://arxiv.org/abs/1107.1716} {arXiv:1107.1716 [astro-ph.CO]} \BibitemShut
  {NoStop}%
\bibitem [{\citenamefont {Smith}\ \emph {et~al.}(2017)\citenamefont {Smith},
  \citenamefont {Muñoz}, \citenamefont {Smith}, \citenamefont {Yee},\ and\
  \citenamefont {Grin}}]{Smith:2017ndr}%
  \BibitemOpen
  \bibfield  {author} {\bibinfo {author} {\bibfnamefont {T.~L.}\ \bibnamefont
  {Smith}}, \bibinfo {author} {\bibfnamefont {J.~B.}\ \bibnamefont {Muñoz}},
  \bibinfo {author} {\bibfnamefont {R.}~\bibnamefont {Smith}}, \bibinfo
  {author} {\bibfnamefont {K.}~\bibnamefont {Yee}}, \ and\ \bibinfo {author}
  {\bibfnamefont {D.}~\bibnamefont {Grin}},\ }\href {\doibase
  10.1103/PhysRevD.96.083508} {\bibfield  {journal} {\bibinfo  {journal} {Phys.
  Rev.}\ }\textbf {\bibinfo {volume} {D96}},\ \bibinfo {pages} {083508}
  (\bibinfo {year} {2017})},\ \Eprint {http://arxiv.org/abs/1704.03461}
  {arXiv:1704.03461 [astro-ph.CO]} \BibitemShut {NoStop}%
\bibitem [{\citenamefont {Muñoz}\ \emph {et~al.}(2016)\citenamefont {Muñoz},
  \citenamefont {Grin}, \citenamefont {Dai}, \citenamefont {Kamionkowski},\
  and\ \citenamefont {Kovetz}}]{Munoz:2015fdv}%
  \BibitemOpen
  \bibfield  {author} {\bibinfo {author} {\bibfnamefont {J.~B.}\ \bibnamefont
  {Muñoz}}, \bibinfo {author} {\bibfnamefont {D.}~\bibnamefont {Grin}},
  \bibinfo {author} {\bibfnamefont {L.}~\bibnamefont {Dai}}, \bibinfo {author}
  {\bibfnamefont {M.}~\bibnamefont {Kamionkowski}}, \ and\ \bibinfo {author}
  {\bibfnamefont {E.~D.}\ \bibnamefont {Kovetz}},\ }\href {\doibase
  10.1103/PhysRevD.93.043008} {\bibfield  {journal} {\bibinfo  {journal} {Phys.
  Rev.}\ }\textbf {\bibinfo {volume} {D93}},\ \bibinfo {pages} {043008}
  (\bibinfo {year} {2016})},\ \Eprint {http://arxiv.org/abs/1511.04441}
  {arXiv:1511.04441 [astro-ph.CO]} \BibitemShut {NoStop}%
\bibitem [{\citenamefont {Grin}\ \emph {et~al.}(2014)\citenamefont {Grin},
  \citenamefont {Hanson}, \citenamefont {Holder}, \citenamefont {Doré},\ and\
  \citenamefont {Kamionkowski}}]{Grin:2013uya}%
  \BibitemOpen
  \bibfield  {author} {\bibinfo {author} {\bibfnamefont {D.}~\bibnamefont
  {Grin}}, \bibinfo {author} {\bibfnamefont {D.}~\bibnamefont {Hanson}},
  \bibinfo {author} {\bibfnamefont {G.~P.}\ \bibnamefont {Holder}}, \bibinfo
  {author} {\bibfnamefont {O.}~\bibnamefont {Doré}}, \ and\ \bibinfo {author}
  {\bibfnamefont {M.}~\bibnamefont {Kamionkowski}},\ }\href {\doibase
  10.1103/PhysRevD.89.023006} {\bibfield  {journal} {\bibinfo  {journal} {Phys.
  Rev.}\ }\textbf {\bibinfo {volume} {D89}},\ \bibinfo {pages} {023006}
  (\bibinfo {year} {2014})},\ \Eprint {http://arxiv.org/abs/1306.4319}
  {arXiv:1306.4319 [astro-ph.CO]} \BibitemShut {NoStop}%
\bibitem [{\citenamefont {Heinrich}\ and\ \citenamefont
  {Schmittfull}(2019)}]{Heinrich:2019sxl}%
  \BibitemOpen
  \bibfield  {author} {\bibinfo {author} {\bibfnamefont {C.}~\bibnamefont
  {Heinrich}}\ and\ \bibinfo {author} {\bibfnamefont {M.}~\bibnamefont
  {Schmittfull}},\ }\href@noop {} {\  (\bibinfo {year} {2019})},\ \Eprint
  {http://arxiv.org/abs/1904.00024} {arXiv:1904.00024 [astro-ph.CO]}
  \BibitemShut {NoStop}%
\bibitem [{\citenamefont {Lyth}\ and\ \citenamefont
  {Wands}(2002)}]{Lyth:2001nq}%
  \BibitemOpen
  \bibfield  {author} {\bibinfo {author} {\bibfnamefont {D.~H.}\ \bibnamefont
  {Lyth}}\ and\ \bibinfo {author} {\bibfnamefont {D.}~\bibnamefont {Wands}},\
  }\href {\doibase 10.1016/S0370-2693(01)01366-1} {\bibfield  {journal}
  {\bibinfo  {journal} {Phys. Lett.}\ }\textbf {\bibinfo {volume} {B524}},\
  \bibinfo {pages} {5} (\bibinfo {year} {2002})},\ \Eprint
  {http://arxiv.org/abs/hep-ph/0110002} {arXiv:hep-ph/0110002 [hep-ph]}
  \BibitemShut {NoStop}%
\bibitem [{\citenamefont {Moroi}\ and\ \citenamefont
  {Takahashi}(2002)}]{Moroi:2002rd}%
  \BibitemOpen
  \bibfield  {author} {\bibinfo {author} {\bibfnamefont {T.}~\bibnamefont
  {Moroi}}\ and\ \bibinfo {author} {\bibfnamefont {T.}~\bibnamefont
  {Takahashi}},\ }\href {\doibase 10.1103/PhysRevD.66.063501} {\bibfield
  {journal} {\bibinfo  {journal} {Phys. Rev.}\ }\textbf {\bibinfo {volume}
  {D66}},\ \bibinfo {pages} {063501} (\bibinfo {year} {2002})},\ \Eprint
  {http://arxiv.org/abs/hep-ph/0206026} {arXiv:hep-ph/0206026 [hep-ph]}
  \BibitemShut {NoStop}%
\bibitem [{\citenamefont {Lyth}\ and\ \citenamefont
  {Wands}(2003)}]{Lyth:2003ip}%
  \BibitemOpen
  \bibfield  {author} {\bibinfo {author} {\bibfnamefont {D.~H.}\ \bibnamefont
  {Lyth}}\ and\ \bibinfo {author} {\bibfnamefont {D.}~\bibnamefont {Wands}},\
  }\href {\doibase 10.1103/PhysRevD.68.103516} {\bibfield  {journal} {\bibinfo
  {journal} {Phys. Rev.}\ }\textbf {\bibinfo {volume} {D68}},\ \bibinfo {pages}
  {103516} (\bibinfo {year} {2003})},\ \Eprint
  {http://arxiv.org/abs/astro-ph/0306500} {arXiv:astro-ph/0306500 [astro-ph]}
  \BibitemShut {NoStop}%
\bibitem [{\citenamefont {{Barkana}}\ and\ \citenamefont
  {{Loeb}}(2011)}]{2011MNRAS.415.3113B}%
  \BibitemOpen
  \bibfield  {author} {\bibinfo {author} {\bibfnamefont {R.}~\bibnamefont
  {{Barkana}}}\ and\ \bibinfo {author} {\bibfnamefont {A.}~\bibnamefont
  {{Loeb}}},\ }\href {\doibase 10.1111/j.1365-2966.2011.18922.x} {\bibfield
  {journal} {\bibinfo  {journal} {\mnras}\ }\textbf {\bibinfo {volume} {415}},\
  \bibinfo {pages} {3113} (\bibinfo {year} {2011})},\ \Eprint
  {http://arxiv.org/abs/1009.1393} {arXiv:1009.1393 [astro-ph.CO]} \BibitemShut
  {NoStop}%
\bibitem [{\citenamefont {Schmidt}(2016)}]{Schmidt:2016coo}%
  \BibitemOpen
  \bibfield  {author} {\bibinfo {author} {\bibfnamefont {F.}~\bibnamefont
  {Schmidt}},\ }\href {\doibase 10.1103/PhysRevD.94.063508} {\bibfield
  {journal} {\bibinfo  {journal} {Phys. Rev.}\ }\textbf {\bibinfo {volume}
  {D94}},\ \bibinfo {pages} {063508} (\bibinfo {year} {2016})},\ \Eprint
  {http://arxiv.org/abs/1602.09059} {arXiv:1602.09059 [astro-ph.CO]}
  \BibitemShut {NoStop}%
\bibitem [{\citenamefont {Barreira}\ \emph {et~al.}(2019)\citenamefont
  {Barreira}, \citenamefont {Cabass}, \citenamefont {Nelson},\ and\
  \citenamefont {Schmidt}}]{Barreira:2019qdl}%
  \BibitemOpen
  \bibfield  {author} {\bibinfo {author} {\bibfnamefont {A.}~\bibnamefont
  {Barreira}}, \bibinfo {author} {\bibfnamefont {G.}~\bibnamefont {Cabass}},
  \bibinfo {author} {\bibfnamefont {D.}~\bibnamefont {Nelson}}, \ and\ \bibinfo
  {author} {\bibfnamefont {F.}~\bibnamefont {Schmidt}},\ }\href@noop {} {\
  (\bibinfo {year} {2019})},\ \Eprint {http://arxiv.org/abs/1907.04317}
  {arXiv:1907.04317 [astro-ph.CO]} \BibitemShut {NoStop}%
\bibitem [{\citenamefont {Dalal}\ \emph {et~al.}(2008)\citenamefont {Dalal},
  \citenamefont {Dore}, \citenamefont {Huterer},\ and\ \citenamefont
  {Shirokov}}]{Dalal:2007cu}%
  \BibitemOpen
  \bibfield  {author} {\bibinfo {author} {\bibfnamefont {N.}~\bibnamefont
  {Dalal}}, \bibinfo {author} {\bibfnamefont {O.}~\bibnamefont {Dore}},
  \bibinfo {author} {\bibfnamefont {D.}~\bibnamefont {Huterer}}, \ and\
  \bibinfo {author} {\bibfnamefont {A.}~\bibnamefont {Shirokov}},\ }\href
  {\doibase 10.1103/PhysRevD.77.123514} {\bibfield  {journal} {\bibinfo
  {journal} {Phys. Rev.}\ }\textbf {\bibinfo {volume} {D77}},\ \bibinfo {pages}
  {123514} (\bibinfo {year} {2008})},\ \Eprint {http://arxiv.org/abs/0710.4560}
  {arXiv:0710.4560 [astro-ph]} \BibitemShut {NoStop}%
\bibitem [{\citenamefont {McDonald}\ and\ \citenamefont
  {Seljak}(2009)}]{McDonald:2008sh}%
  \BibitemOpen
  \bibfield  {author} {\bibinfo {author} {\bibfnamefont {P.}~\bibnamefont
  {McDonald}}\ and\ \bibinfo {author} {\bibfnamefont {U.}~\bibnamefont
  {Seljak}},\ }\href {\doibase 10.1088/1475-7516/2009/10/007} {\bibfield
  {journal} {\bibinfo  {journal} {JCAP}\ }\textbf {\bibinfo {volume} {0910}},\
  \bibinfo {pages} {007} (\bibinfo {year} {2009})},\ \Eprint
  {http://arxiv.org/abs/0810.0323} {arXiv:0810.0323 [astro-ph]} \BibitemShut
  {NoStop}%
\bibitem [{\citenamefont {Seljak}(2009)}]{Seljak:2008xr}%
  \BibitemOpen
  \bibfield  {author} {\bibinfo {author} {\bibfnamefont {U.}~\bibnamefont
  {Seljak}},\ }\href {\doibase 10.1103/PhysRevLett.102.021302} {\bibfield
  {journal} {\bibinfo  {journal} {Phys. Rev. Lett.}\ }\textbf {\bibinfo
  {volume} {102}},\ \bibinfo {pages} {021302} (\bibinfo {year} {2009})},\
  \Eprint {http://arxiv.org/abs/0807.1770} {arXiv:0807.1770 [astro-ph]}
  \BibitemShut {NoStop}%
\bibitem [{\citenamefont {M{\"u}nchmeyer}\ \emph {et~al.}(2018)\citenamefont
  {M{\"u}nchmeyer}, \citenamefont {Madhavacheril}, \citenamefont {Ferraro},
  \citenamefont {Johnson},\ and\ \citenamefont {Smith}}]{Munchmeyer:2018eey}%
  \BibitemOpen
  \bibfield  {author} {\bibinfo {author} {\bibfnamefont {M.}~\bibnamefont
  {M{\"u}nchmeyer}}, \bibinfo {author} {\bibfnamefont {M.~S.}\ \bibnamefont
  {Madhavacheril}}, \bibinfo {author} {\bibfnamefont {S.}~\bibnamefont
  {Ferraro}}, \bibinfo {author} {\bibfnamefont {M.~C.}\ \bibnamefont
  {Johnson}}, \ and\ \bibinfo {author} {\bibfnamefont {K.~M.}\ \bibnamefont
  {Smith}},\ }\href@noop {} {\  (\bibinfo {year} {2018})},\ \Eprint
  {http://arxiv.org/abs/1810.13424} {arXiv:1810.13424 [astro-ph.CO]}
  \BibitemShut {NoStop}%
\bibitem [{\citenamefont {Contreras}\ \emph {et~al.}(2019)\citenamefont
  {Contreras}, \citenamefont {Johnson},\ and\ \citenamefont
  {Mertens}}]{Contreras:2019bxy}%
  \BibitemOpen
  \bibfield  {author} {\bibinfo {author} {\bibfnamefont {D.}~\bibnamefont
  {Contreras}}, \bibinfo {author} {\bibfnamefont {M.~C.}\ \bibnamefont
  {Johnson}}, \ and\ \bibinfo {author} {\bibfnamefont {J.~B.}\ \bibnamefont
  {Mertens}},\ }\href@noop {} {\  (\bibinfo {year} {2019})},\ \Eprint
  {http://arxiv.org/abs/1904.10033} {arXiv:1904.10033 [astro-ph.CO]}
  \BibitemShut {NoStop}%
\bibitem [{\citenamefont {{Zhang}}(2010)}]{Zhang10d}%
  \BibitemOpen
  \bibfield  {author} {\bibinfo {author} {\bibfnamefont {P.}~\bibnamefont
  {{Zhang}}},\ }\href {\doibase 10.1111/j.1745-3933.2010.00899.x} {\bibfield
  {journal} {\bibinfo  {journal} {MNRAS}\ }\textbf {\bibinfo {volume} {407}},\
  \bibinfo {pages} {L36} (\bibinfo {year} {2010})},\ \Eprint
  {http://arxiv.org/abs/1004.0990} {arXiv:1004.0990 [astro-ph.CO]} \BibitemShut
  {NoStop}%
\bibitem [{\citenamefont {Zhang}\ and\ \citenamefont
  {Johnson}(2015)}]{Zhang:2015uta}%
  \BibitemOpen
  \bibfield  {author} {\bibinfo {author} {\bibfnamefont {P.}~\bibnamefont
  {Zhang}}\ and\ \bibinfo {author} {\bibfnamefont {M.~C.}\ \bibnamefont
  {Johnson}},\ }\href {\doibase 10.1088/1475-7516/2015/06/046} {\bibfield
  {journal} {\bibinfo  {journal} {JCAP}\ }\textbf {\bibinfo {volume} {1506}},\
  \bibinfo {pages} {046} (\bibinfo {year} {2015})},\ \Eprint
  {http://arxiv.org/abs/1501.00511} {arXiv:1501.00511 [astro-ph.CO]}
  \BibitemShut {NoStop}%
\bibitem [{\citenamefont {Terrana}\ \emph {et~al.}(2017)\citenamefont
  {Terrana}, \citenamefont {Harris},\ and\ \citenamefont
  {Johnson}}]{Terrana:2016xvc}%
  \BibitemOpen
  \bibfield  {author} {\bibinfo {author} {\bibfnamefont {A.}~\bibnamefont
  {Terrana}}, \bibinfo {author} {\bibfnamefont {M.-J.}\ \bibnamefont {Harris}},
  \ and\ \bibinfo {author} {\bibfnamefont {M.~C.}\ \bibnamefont {Johnson}},\
  }\href {\doibase 10.1088/1475-7516/2017/02/040} {\bibfield  {journal}
  {\bibinfo  {journal} {JCAP}\ }\textbf {\bibinfo {volume} {1702}},\ \bibinfo
  {pages} {040} (\bibinfo {year} {2017})},\ \Eprint
  {http://arxiv.org/abs/1610.06919} {arXiv:1610.06919 [astro-ph.CO]}
  \BibitemShut {NoStop}%
\bibitem [{\citenamefont {Deutsch}\ \emph
  {et~al.}(2018{\natexlab{a}})\citenamefont {Deutsch}, \citenamefont
  {Dimastrogiovanni}, \citenamefont {Johnson}, \citenamefont {Münchmeyer},\
  and\ \citenamefont {Terrana}}]{Deutsch:2017ybc}%
  \BibitemOpen
  \bibfield  {author} {\bibinfo {author} {\bibfnamefont {A.-S.}\ \bibnamefont
  {Deutsch}}, \bibinfo {author} {\bibfnamefont {E.}~\bibnamefont
  {Dimastrogiovanni}}, \bibinfo {author} {\bibfnamefont {M.~C.}\ \bibnamefont
  {Johnson}}, \bibinfo {author} {\bibfnamefont {M.}~\bibnamefont
  {Münchmeyer}}, \ and\ \bibinfo {author} {\bibfnamefont {A.}~\bibnamefont
  {Terrana}},\ }\href {\doibase 10.1103/PhysRevD.98.123501} {\bibfield
  {journal} {\bibinfo  {journal} {Phys. Rev.}\ }\textbf {\bibinfo {volume}
  {D98}},\ \bibinfo {pages} {123501} (\bibinfo {year} {2018}{\natexlab{a}})},\
  \Eprint {http://arxiv.org/abs/1707.08129} {arXiv:1707.08129 [astro-ph.CO]}
  \BibitemShut {NoStop}%
\bibitem [{\citenamefont {Smith}\ \emph {et~al.}(2018)\citenamefont {Smith},
  \citenamefont {Madhavacheril}, \citenamefont {M{\"u}nchmeyer}, \citenamefont
  {Ferraro}, \citenamefont {Giri},\ and\ \citenamefont
  {Johnson}}]{Smith:2018bpn}%
  \BibitemOpen
  \bibfield  {author} {\bibinfo {author} {\bibfnamefont {K.~M.}\ \bibnamefont
  {Smith}}, \bibinfo {author} {\bibfnamefont {M.~S.}\ \bibnamefont
  {Madhavacheril}}, \bibinfo {author} {\bibfnamefont {M.}~\bibnamefont
  {M{\"u}nchmeyer}}, \bibinfo {author} {\bibfnamefont {S.}~\bibnamefont
  {Ferraro}}, \bibinfo {author} {\bibfnamefont {U.}~\bibnamefont {Giri}}, \
  and\ \bibinfo {author} {\bibfnamefont {M.~C.}\ \bibnamefont {Johnson}},\
  }\href@noop {} {\  (\bibinfo {year} {2018})},\ \Eprint
  {http://arxiv.org/abs/1810.13423} {arXiv:1810.13423 [astro-ph.CO]}
  \BibitemShut {NoStop}%
\bibitem [{\citenamefont {Aguirre}\ \emph {et~al.}(2018)\citenamefont {Aguirre}
  \emph {et~al.}}]{Ade:2018sbj}%
  \BibitemOpen
  \bibfield  {author} {\bibinfo {author} {\bibfnamefont {J.}~\bibnamefont
  {Aguirre}} \emph {et~al.} (\bibinfo {collaboration} {Simons Observatory}),\
  }\href@noop {} {\  (\bibinfo {year} {2018})},\ \Eprint
  {http://arxiv.org/abs/1808.07445} {arXiv:1808.07445 [astro-ph.CO]}
  \BibitemShut {NoStop}%
\bibitem [{\citenamefont {Abazajian}\ \emph {et~al.}(2016)\citenamefont
  {Abazajian} \emph {et~al.}}]{Abazajian:2016yjj}%
  \BibitemOpen
  \bibfield  {author} {\bibinfo {author} {\bibfnamefont {K.~N.}\ \bibnamefont
  {Abazajian}} \emph {et~al.} (\bibinfo {collaboration} {CMB-S4}),\ }\href@noop
  {} {\  (\bibinfo {year} {2016})},\ \Eprint {http://arxiv.org/abs/1610.02743}
  {arXiv:1610.02743 [astro-ph.CO]} \BibitemShut {NoStop}%
\bibitem [{\citenamefont {Abell}\ \emph {et~al.}(2009)\citenamefont {Abell}
  \emph {et~al.}}]{Abell:2009aa}%
  \BibitemOpen
  \bibfield  {author} {\bibinfo {author} {\bibfnamefont {P.~A.}\ \bibnamefont
  {Abell}} \emph {et~al.} (\bibinfo {collaboration} {LSST Science, LSST
  Project}),\ }\href@noop {} {\  (\bibinfo {year} {2009})},\ \Eprint
  {http://arxiv.org/abs/0912.0201} {arXiv:0912.0201 [astro-ph.IM]} \BibitemShut
  {NoStop}%
\bibitem [{\citenamefont {Aghamousa}\ \emph {et~al.}(2016)\citenamefont
  {Aghamousa} \emph {et~al.}}]{Aghamousa:2016zmz}%
  \BibitemOpen
  \bibfield  {author} {\bibinfo {author} {\bibfnamefont {A.}~\bibnamefont
  {Aghamousa}} \emph {et~al.} (\bibinfo {collaboration} {DESI}),\ }\href@noop
  {} {\  (\bibinfo {year} {2016})},\ \Eprint {http://arxiv.org/abs/1611.00036}
  {arXiv:1611.00036 [astro-ph.IM]} \BibitemShut {NoStop}%
\bibitem [{\citenamefont {Pan}\ and\ \citenamefont
  {Johnson}(2019)}]{Pan:2019dax}%
  \BibitemOpen
  \bibfield  {author} {\bibinfo {author} {\bibfnamefont {Z.}~\bibnamefont
  {Pan}}\ and\ \bibinfo {author} {\bibfnamefont {M.~C.}\ \bibnamefont
  {Johnson}},\ }\href@noop {} {\  (\bibinfo {year} {2019})},\ \Eprint
  {http://arxiv.org/abs/1906.04208} {arXiv:1906.04208 [astro-ph.CO]}
  \BibitemShut {NoStop}%
\bibitem [{\citenamefont {Cayuso}\ and\ \citenamefont
  {Johnson}(2019)}]{Cayuso:2019hen}%
  \BibitemOpen
  \bibfield  {author} {\bibinfo {author} {\bibfnamefont {J.~I.}\ \bibnamefont
  {Cayuso}}\ and\ \bibinfo {author} {\bibfnamefont {M.~C.}\ \bibnamefont
  {Johnson}},\ }\href@noop {} {\  (\bibinfo {year} {2019})},\ \Eprint
  {http://arxiv.org/abs/1904.10981} {arXiv:1904.10981 [astro-ph.CO]}
  \BibitemShut {NoStop}%
\bibitem [{\citenamefont {He}\ \emph {et~al.}(2015)\citenamefont {He},
  \citenamefont {Grin},\ and\ \citenamefont {Hu}}]{He:2015msa}%
  \BibitemOpen
  \bibfield  {author} {\bibinfo {author} {\bibfnamefont {C.}~\bibnamefont
  {He}}, \bibinfo {author} {\bibfnamefont {D.}~\bibnamefont {Grin}}, \ and\
  \bibinfo {author} {\bibfnamefont {W.}~\bibnamefont {Hu}},\ }\href {\doibase
  10.1103/PhysRevD.92.063018} {\bibfield  {journal} {\bibinfo  {journal} {Phys.
  Rev.}\ }\textbf {\bibinfo {volume} {D92}},\ \bibinfo {pages} {063018}
  (\bibinfo {year} {2015})},\ \Eprint {http://arxiv.org/abs/1505.00639}
  {arXiv:1505.00639 [astro-ph.CO]} \BibitemShut {NoStop}%
\bibitem [{\citenamefont {Linde}\ and\ \citenamefont
  {Mukhanov}(1997)}]{Linde:1996gt}%
  \BibitemOpen
  \bibfield  {author} {\bibinfo {author} {\bibfnamefont {A.~D.}\ \bibnamefont
  {Linde}}\ and\ \bibinfo {author} {\bibfnamefont {V.~F.}\ \bibnamefont
  {Mukhanov}},\ }\href {\doibase 10.1103/PhysRevD.56.R535} {\bibfield
  {journal} {\bibinfo  {journal} {Phys. Rev.}\ }\textbf {\bibinfo {volume}
  {D56}},\ \bibinfo {pages} {R535} (\bibinfo {year} {1997})},\ \Eprint
  {http://arxiv.org/abs/astro-ph/9610219} {arXiv:astro-ph/9610219 [astro-ph]}
  \BibitemShut {NoStop}%
\bibitem [{\citenamefont {Moroi}\ and\ \citenamefont
  {Takahashi}(2001)}]{Moroi:2001ct}%
  \BibitemOpen
  \bibfield  {author} {\bibinfo {author} {\bibfnamefont {T.}~\bibnamefont
  {Moroi}}\ and\ \bibinfo {author} {\bibfnamefont {T.}~\bibnamefont
  {Takahashi}},\ }\href {\doibase 10.1016/S0370-2693(02)02070-1,
  10.1016/S0370-2693(01)01295-3} {\bibfield  {journal} {\bibinfo  {journal}
  {Phys. Lett.}\ }\textbf {\bibinfo {volume} {B522}},\ \bibinfo {pages} {215}
  (\bibinfo {year} {2001})},\ \bibinfo {note} {[Erratum: Phys.
  Lett.B539,303(2002)]},\ \Eprint {http://arxiv.org/abs/hep-ph/0110096}
  {arXiv:hep-ph/0110096 [hep-ph]} \BibitemShut {NoStop}%
\bibitem [{\citenamefont {Gordon}\ and\ \citenamefont
  {Lewis}(2003)}]{Gordon:2002gv}%
  \BibitemOpen
  \bibfield  {author} {\bibinfo {author} {\bibfnamefont {C.}~\bibnamefont
  {Gordon}}\ and\ \bibinfo {author} {\bibfnamefont {A.}~\bibnamefont {Lewis}},\
  }\href {\doibase 10.1103/PhysRevD.67.123513} {\bibfield  {journal} {\bibinfo
  {journal} {Phys. Rev.}\ }\textbf {\bibinfo {volume} {D67}},\ \bibinfo {pages}
  {123513} (\bibinfo {year} {2003})},\ \Eprint
  {http://arxiv.org/abs/astro-ph/0212248} {arXiv:astro-ph/0212248 [astro-ph]}
  \BibitemShut {NoStop}%
\bibitem [{\citenamefont {Heinrich}\ \emph {et~al.}(2016)\citenamefont
  {Heinrich}, \citenamefont {Grin},\ and\ \citenamefont
  {Hu}}]{Heinrich:2016gqe}%
  \BibitemOpen
  \bibfield  {author} {\bibinfo {author} {\bibfnamefont {C.~H.}\ \bibnamefont
  {Heinrich}}, \bibinfo {author} {\bibfnamefont {D.}~\bibnamefont {Grin}}, \
  and\ \bibinfo {author} {\bibfnamefont {W.}~\bibnamefont {Hu}},\ }\href
  {\doibase 10.1103/PhysRevD.94.043534} {\bibfield  {journal} {\bibinfo
  {journal} {Phys. Rev.}\ }\textbf {\bibinfo {volume} {D94}},\ \bibinfo {pages}
  {043534} (\bibinfo {year} {2016})},\ \Eprint
  {http://arxiv.org/abs/1605.08439} {arXiv:1605.08439 [astro-ph.CO]}
  \BibitemShut {NoStop}%
\bibitem [{\citenamefont {Sasaki}\ \emph {et~al.}(2006)\citenamefont {Sasaki},
  \citenamefont {Valiviita},\ and\ \citenamefont {Wands}}]{Sasaki:2006kq}%
  \BibitemOpen
  \bibfield  {author} {\bibinfo {author} {\bibfnamefont {M.}~\bibnamefont
  {Sasaki}}, \bibinfo {author} {\bibfnamefont {J.}~\bibnamefont {Valiviita}}, \
  and\ \bibinfo {author} {\bibfnamefont {D.}~\bibnamefont {Wands}},\ }\href
  {\doibase 10.1103/PhysRevD.74.103003} {\bibfield  {journal} {\bibinfo
  {journal} {Phys. Rev.}\ }\textbf {\bibinfo {volume} {D74}},\ \bibinfo {pages}
  {103003} (\bibinfo {year} {2006})},\ \Eprint
  {http://arxiv.org/abs/astro-ph/0607627} {arXiv:astro-ph/0607627 [astro-ph]}
  \BibitemShut {NoStop}%
\bibitem [{\citenamefont {{Bonvin}}\ and\ \citenamefont
  {{Durrer}}(2011)}]{2011PhRvD..84f3505B}%
  \BibitemOpen
  \bibfield  {author} {\bibinfo {author} {\bibfnamefont {C.}~\bibnamefont
  {{Bonvin}}}\ and\ \bibinfo {author} {\bibfnamefont {R.}~\bibnamefont
  {{Durrer}}},\ }\href {\doibase 10.1103/PhysRevD.84.063505} {\bibfield
  {journal} {\bibinfo  {journal} {\prd}\ }\textbf {\bibinfo {volume} {84}},\
  \bibinfo {eid} {063505} (\bibinfo {year} {2011})},\ \Eprint
  {http://arxiv.org/abs/1105.5280} {arXiv:1105.5280 [astro-ph.CO]} \BibitemShut
  {NoStop}%
\bibitem [{\citenamefont {{Challinor}}\ and\ \citenamefont
  {{Lewis}}(2011)}]{2011PhRvD..84d3516C}%
  \BibitemOpen
  \bibfield  {author} {\bibinfo {author} {\bibfnamefont {A.}~\bibnamefont
  {{Challinor}}}\ and\ \bibinfo {author} {\bibfnamefont {A.}~\bibnamefont
  {{Lewis}}},\ }\href {\doibase 10.1103/PhysRevD.84.043516} {\bibfield
  {journal} {\bibinfo  {journal} {\prd}\ }\textbf {\bibinfo {volume} {84}},\
  \bibinfo {eid} {043516} (\bibinfo {year} {2011})},\ \Eprint
  {http://arxiv.org/abs/1105.5292} {arXiv:1105.5292 [astro-ph.CO]} \BibitemShut
  {NoStop}%
\bibitem [{\citenamefont {Hirata}(2009)}]{0903.4929}%
  \BibitemOpen
  \bibfield  {author} {\bibinfo {author} {\bibfnamefont {C.~M.}\ \bibnamefont
  {Hirata}},\ }\href {\doibase 10.1111/j.1365-2966.2009.15353.x} {\bibfield
  {journal} {\bibinfo  {journal} {Mon. Not. Roy. Astron. Soc.}\ }\textbf
  {\bibinfo {volume} {399}},\ \bibinfo {pages} {1074} (\bibinfo {year}
  {2009})},\ \Eprint {http://arxiv.org/abs/0903.4929} {arXiv:0903.4929
  [astro-ph.CO]} \BibitemShut {NoStop}%
\bibitem [{\citenamefont {Madhavacheril}\ \emph {et~al.}(2019)\citenamefont
  {Madhavacheril}, \citenamefont {Battaglia}, \citenamefont {Smith},\ and\
  \citenamefont {Sievers}}]{Madhavacheril:2019buy}%
  \BibitemOpen
  \bibfield  {author} {\bibinfo {author} {\bibfnamefont {M.~S.}\ \bibnamefont
  {Madhavacheril}}, \bibinfo {author} {\bibfnamefont {N.}~\bibnamefont
  {Battaglia}}, \bibinfo {author} {\bibfnamefont {K.~M.}\ \bibnamefont
  {Smith}}, \ and\ \bibinfo {author} {\bibfnamefont {J.~L.}\ \bibnamefont
  {Sievers}},\ }\href@noop {} {\  (\bibinfo {year} {2019})},\ \Eprint
  {http://arxiv.org/abs/1901.02418} {arXiv:1901.02418 [astro-ph.CO]}
  \BibitemShut {NoStop}%
\bibitem [{\citenamefont {Mueller}\ \emph {et~al.}(2015)\citenamefont
  {Mueller}, \citenamefont {de~Bernardis}, \citenamefont {Bean},\ and\
  \citenamefont {Niemack}}]{Mueller:2014nsa}%
  \BibitemOpen
  \bibfield  {author} {\bibinfo {author} {\bibfnamefont {E.-M.}\ \bibnamefont
  {Mueller}}, \bibinfo {author} {\bibfnamefont {F.}~\bibnamefont
  {de~Bernardis}}, \bibinfo {author} {\bibfnamefont {R.}~\bibnamefont {Bean}},
  \ and\ \bibinfo {author} {\bibfnamefont {M.~D.}\ \bibnamefont {Niemack}},\
  }\href {\doibase 10.1088/0004-637X/808/1/47} {\bibfield  {journal} {\bibinfo
  {journal} {Astrophys. J.}\ }\textbf {\bibinfo {volume} {808}},\ \bibinfo
  {pages} {47} (\bibinfo {year} {2015})},\ \Eprint
  {http://arxiv.org/abs/1408.6248} {arXiv:1408.6248 [astro-ph.CO]} \BibitemShut
  {NoStop}%
\bibitem [{\citenamefont {{Battaglia}}(2016)}]{Battaglia2016}%
  \BibitemOpen
  \bibfield  {author} {\bibinfo {author} {\bibfnamefont {N.}~\bibnamefont
  {{Battaglia}}},\ }\href {\doibase 10.1088/1475-7516/2016/08/058} {\bibfield
  {journal} {\bibinfo  {journal} {\jcap}\ }\textbf {\bibinfo {volume} {8}},\
  \bibinfo {eid} {058} (\bibinfo {year} {2016})},\ \Eprint
  {http://arxiv.org/abs/1607.02442} {arXiv:1607.02442} \BibitemShut {NoStop}%
\bibitem [{\citenamefont {Lorenz}\ \emph {et~al.}(2018)\citenamefont {Lorenz},
  \citenamefont {Alonso},\ and\ \citenamefont {Ferreira}}]{Lorenz:2017iez}%
  \BibitemOpen
  \bibfield  {author} {\bibinfo {author} {\bibfnamefont {C.~S.}\ \bibnamefont
  {Lorenz}}, \bibinfo {author} {\bibfnamefont {D.}~\bibnamefont {Alonso}}, \
  and\ \bibinfo {author} {\bibfnamefont {P.~G.}\ \bibnamefont {Ferreira}},\
  }\href {\doibase 10.1103/PhysRevD.97.023537} {\bibfield  {journal} {\bibinfo
  {journal} {Phys. Rev.}\ }\textbf {\bibinfo {volume} {D97}},\ \bibinfo {pages}
  {023537} (\bibinfo {year} {2018})},\ \Eprint
  {http://arxiv.org/abs/1710.02477} {arXiv:1710.02477 [astro-ph.CO]}
  \BibitemShut {NoStop}%
\bibitem [{\citenamefont {Alonso}\ \emph {et~al.}(2015)\citenamefont {Alonso},
  \citenamefont {Bull}, \citenamefont {Ferreira}, \citenamefont {Maartens},\
  and\ \citenamefont {Santos}}]{Alonso:2015uua}%
  \BibitemOpen
  \bibfield  {author} {\bibinfo {author} {\bibfnamefont {D.}~\bibnamefont
  {Alonso}}, \bibinfo {author} {\bibfnamefont {P.}~\bibnamefont {Bull}},
  \bibinfo {author} {\bibfnamefont {P.~G.}\ \bibnamefont {Ferreira}}, \bibinfo
  {author} {\bibfnamefont {R.}~\bibnamefont {Maartens}}, \ and\ \bibinfo
  {author} {\bibfnamefont {M.}~\bibnamefont {Santos}},\ }\href {\doibase
  10.1088/0004-637X/814/2/145} {\bibfield  {journal} {\bibinfo  {journal}
  {Astrophys. J.}\ }\textbf {\bibinfo {volume} {814}},\ \bibinfo {pages} {145}
  (\bibinfo {year} {2015})},\ \Eprint {http://arxiv.org/abs/1505.07596}
  {arXiv:1505.07596 [astro-ph.CO]} \BibitemShut {NoStop}%
\bibitem [{\citenamefont {Weaverdyck}\ \emph {et~al.}(2018)\citenamefont
  {Weaverdyck}, \citenamefont {Muir},\ and\ \citenamefont
  {Huterer}}]{Weaverdyck:2017ovf}%
  \BibitemOpen
  \bibfield  {author} {\bibinfo {author} {\bibfnamefont {N.}~\bibnamefont
  {Weaverdyck}}, \bibinfo {author} {\bibfnamefont {J.}~\bibnamefont {Muir}}, \
  and\ \bibinfo {author} {\bibfnamefont {D.}~\bibnamefont {Huterer}},\ }\href
  {\doibase 10.1103/PhysRevD.97.043515} {\bibfield  {journal} {\bibinfo
  {journal} {Phys. Rev.}\ }\textbf {\bibinfo {volume} {D97}},\ \bibinfo {pages}
  {043515} (\bibinfo {year} {2018})},\ \Eprint
  {http://arxiv.org/abs/1709.08661} {arXiv:1709.08661 [astro-ph.CO]}
  \BibitemShut {NoStop}%
\bibitem [{\citenamefont {Sunyaev}\ and\ \citenamefont
  {Zeldovich}(1980)}]{10.1093/mnras/190.3.413}%
  \BibitemOpen
  \bibfield  {author} {\bibinfo {author} {\bibfnamefont {R.~A.}\ \bibnamefont
  {Sunyaev}}\ and\ \bibinfo {author} {\bibfnamefont {Y.~B.}\ \bibnamefont
  {Zeldovich}},\ }\href {\doibase 10.1093/mnras/190.3.413} {\bibfield
  {journal} {\bibinfo  {journal} {Monthly Notices of the Royal Astronomical
  Society}\ }\textbf {\bibinfo {volume} {190}},\ \bibinfo {pages} {413}
  (\bibinfo {year} {1980})}\BibitemShut {NoStop}%
\bibitem [{\citenamefont {Kamionkowski}\ and\ \citenamefont
  {Loeb}(1997)}]{Kamionkowski:1997na}%
  \BibitemOpen
  \bibfield  {author} {\bibinfo {author} {\bibfnamefont {M.}~\bibnamefont
  {Kamionkowski}}\ and\ \bibinfo {author} {\bibfnamefont {A.}~\bibnamefont
  {Loeb}},\ }\href {\doibase 10.1103/PhysRevD.56.4511} {\bibfield  {journal}
  {\bibinfo  {journal} {Phys. Rev.}\ }\textbf {\bibinfo {volume} {D56}},\
  \bibinfo {pages} {4511} (\bibinfo {year} {1997})},\ \Eprint
  {http://arxiv.org/abs/astro-ph/9703118} {arXiv:astro-ph/9703118 [astro-ph]}
  \BibitemShut {NoStop}%
\bibitem [{\citenamefont {Deutsch}\ \emph
  {et~al.}(2018{\natexlab{b}})\citenamefont {Deutsch}, \citenamefont {Johnson},
  \citenamefont {Münchmeyer},\ and\ \citenamefont
  {Terrana}}]{Deutsch:2017cja}%
  \BibitemOpen
  \bibfield  {author} {\bibinfo {author} {\bibfnamefont {A.-S.}\ \bibnamefont
  {Deutsch}}, \bibinfo {author} {\bibfnamefont {M.~C.}\ \bibnamefont
  {Johnson}}, \bibinfo {author} {\bibfnamefont {M.}~\bibnamefont
  {Münchmeyer}}, \ and\ \bibinfo {author} {\bibfnamefont {A.}~\bibnamefont
  {Terrana}},\ }\href {\doibase 10.1088/1475-7516/2018/04/034} {\bibfield
  {journal} {\bibinfo  {journal} {JCAP}\ }\textbf {\bibinfo {volume} {1804}},\
  \bibinfo {pages} {034} (\bibinfo {year} {2018}{\natexlab{b}})},\ \Eprint
  {http://arxiv.org/abs/1705.08907} {arXiv:1705.08907 [astro-ph.CO]}
  \BibitemShut {NoStop}%
\bibitem [{\citenamefont {Meyers}\ \emph {et~al.}(2018)\citenamefont {Meyers},
  \citenamefont {Meerburg}, \citenamefont {van Engelen},\ and\ \citenamefont
  {Battaglia}}]{Meyers:2017rtf}%
  \BibitemOpen
  \bibfield  {author} {\bibinfo {author} {\bibfnamefont {J.}~\bibnamefont
  {Meyers}}, \bibinfo {author} {\bibfnamefont {P.~D.}\ \bibnamefont
  {Meerburg}}, \bibinfo {author} {\bibfnamefont {A.}~\bibnamefont {van
  Engelen}}, \ and\ \bibinfo {author} {\bibfnamefont {N.}~\bibnamefont
  {Battaglia}},\ }\href {\doibase 10.1103/PhysRevD.97.103505} {\bibfield
  {journal} {\bibinfo  {journal} {Phys. Rev.}\ }\textbf {\bibinfo {volume}
  {D97}},\ \bibinfo {pages} {103505} (\bibinfo {year} {2018})},\ \Eprint
  {http://arxiv.org/abs/1710.01708} {arXiv:1710.01708 [astro-ph.CO]}
  \BibitemShut {NoStop}%
\bibitem [{\citenamefont {{Birkinshaw}}\ and\ \citenamefont
  {{Gull}}(1983)}]{1983Natur.302..315B}%
  \BibitemOpen
  \bibfield  {author} {\bibinfo {author} {\bibfnamefont {M.}~\bibnamefont
  {{Birkinshaw}}}\ and\ \bibinfo {author} {\bibfnamefont {S.~F.}\ \bibnamefont
  {{Gull}}},\ }\href {\doibase 10.1038/302315a0} {\bibfield  {journal}
  {\bibinfo  {journal} {\nat}\ }\textbf {\bibinfo {volume} {302}},\ \bibinfo
  {pages} {315} (\bibinfo {year} {1983})}\BibitemShut {NoStop}%
\bibitem [{\citenamefont {Hotinli}\ \emph {et~al.}(2019)\citenamefont
  {Hotinli}, \citenamefont {Meyers}, \citenamefont {Dalal}, \citenamefont
  {Jaffe}, \citenamefont {Johnson}, \citenamefont {Mertens}, \citenamefont
  {Münchmeyer}, \citenamefont {Smith},\ and\ \citenamefont {van
  Engelen}}]{Hotinli:2018yyc}%
  \BibitemOpen
  \bibfield  {author} {\bibinfo {author} {\bibfnamefont {S.~C.}\ \bibnamefont
  {Hotinli}}, \bibinfo {author} {\bibfnamefont {J.}~\bibnamefont {Meyers}},
  \bibinfo {author} {\bibfnamefont {N.}~\bibnamefont {Dalal}}, \bibinfo
  {author} {\bibfnamefont {A.~H.}\ \bibnamefont {Jaffe}}, \bibinfo {author}
  {\bibfnamefont {M.~C.}\ \bibnamefont {Johnson}}, \bibinfo {author}
  {\bibfnamefont {J.~B.}\ \bibnamefont {Mertens}}, \bibinfo {author}
  {\bibfnamefont {M.}~\bibnamefont {Münchmeyer}}, \bibinfo {author}
  {\bibfnamefont {K.~M.}\ \bibnamefont {Smith}}, \ and\ \bibinfo {author}
  {\bibfnamefont {A.}~\bibnamefont {van Engelen}},\ }\href {\doibase
  10.1103/PhysRevLett.123.061301} {\bibfield  {journal} {\bibinfo  {journal}
  {Phys. Rev. Lett.}\ }\textbf {\bibinfo {volume} {123}},\ \bibinfo {pages}
  {061301} (\bibinfo {year} {2019})},\ \Eprint
  {http://arxiv.org/abs/1812.03167} {arXiv:1812.03167 [astro-ph.CO]}
  \BibitemShut {NoStop}%
\bibitem [{\citenamefont {Yasini}\ \emph {et~al.}(2019)\citenamefont {Yasini},
  \citenamefont {Mirzatuny},\ and\ \citenamefont {Pierpaoli}}]{Yasini:2018rrl}%
  \BibitemOpen
  \bibfield  {author} {\bibinfo {author} {\bibfnamefont {S.}~\bibnamefont
  {Yasini}}, \bibinfo {author} {\bibfnamefont {N.}~\bibnamefont {Mirzatuny}}, \
  and\ \bibinfo {author} {\bibfnamefont {E.}~\bibnamefont {Pierpaoli}},\ }\href
  {\doibase 10.3847/2041-8213/ab0bfe} {\bibfield  {journal} {\bibinfo
  {journal} {Astrophys. J.}\ }\textbf {\bibinfo {volume} {873}},\ \bibinfo
  {pages} {L23} (\bibinfo {year} {2019})},\ \Eprint
  {http://arxiv.org/abs/1812.04241} {arXiv:1812.04241 [astro-ph.CO]}
  \BibitemShut {NoStop}%
\bibitem [{\citenamefont {Baumann}\ \emph {et~al.}(2013)\citenamefont
  {Baumann}, \citenamefont {Ferraro}, \citenamefont {Green},\ and\
  \citenamefont {Smith}}]{Baumann:2012bc}%
  \BibitemOpen
  \bibfield  {author} {\bibinfo {author} {\bibfnamefont {D.}~\bibnamefont
  {Baumann}}, \bibinfo {author} {\bibfnamefont {S.}~\bibnamefont {Ferraro}},
  \bibinfo {author} {\bibfnamefont {D.}~\bibnamefont {Green}}, \ and\ \bibinfo
  {author} {\bibfnamefont {K.~M.}\ \bibnamefont {Smith}},\ }\href {\doibase
  10.1088/1475-7516/2013/05/001} {\bibfield  {journal} {\bibinfo  {journal}
  {JCAP}\ }\textbf {\bibinfo {volume} {1305}},\ \bibinfo {pages} {001}
  (\bibinfo {year} {2013})},\ \Eprint {http://arxiv.org/abs/1209.2173}
  {arXiv:1209.2173 [astro-ph.CO]} \BibitemShut {NoStop}%
\bibitem [{\citenamefont {Leistedt}\ \emph {et~al.}(2014)\citenamefont
  {Leistedt}, \citenamefont {Peiris},\ and\ \citenamefont
  {Roth}}]{Leistedt:2014zqa}%
  \BibitemOpen
  \bibfield  {author} {\bibinfo {author} {\bibfnamefont {B.}~\bibnamefont
  {Leistedt}}, \bibinfo {author} {\bibfnamefont {H.~V.}\ \bibnamefont
  {Peiris}}, \ and\ \bibinfo {author} {\bibfnamefont {N.}~\bibnamefont
  {Roth}},\ }\href {\doibase 10.1103/PhysRevLett.113.221301} {\bibfield
  {journal} {\bibinfo  {journal} {Phys. Rev. Lett.}\ }\textbf {\bibinfo
  {volume} {113}},\ \bibinfo {pages} {221301} (\bibinfo {year} {2014})},\
  \Eprint {http://arxiv.org/abs/1405.4315} {arXiv:1405.4315 [astro-ph.CO]}
  \BibitemShut {NoStop}%
\bibitem [{\citenamefont {Raveri}\ \emph {et~al.}(2016)\citenamefont {Raveri},
  \citenamefont {Martinelli}, \citenamefont {Zhao},\ and\ \citenamefont
  {Wang}}]{Raveri:2016leq}%
  \BibitemOpen
  \bibfield  {author} {\bibinfo {author} {\bibfnamefont {M.}~\bibnamefont
  {Raveri}}, \bibinfo {author} {\bibfnamefont {M.}~\bibnamefont {Martinelli}},
  \bibinfo {author} {\bibfnamefont {G.}~\bibnamefont {Zhao}}, \ and\ \bibinfo
  {author} {\bibfnamefont {Y.}~\bibnamefont {Wang}},\ }\href@noop {} {\
  (\bibinfo {year} {2016})},\ \Eprint {http://arxiv.org/abs/1606.06268}
  {arXiv:1606.06268 [astro-ph.CO]} \BibitemShut {NoStop}%
\end{thebibliography}%

\end{document}